\documentclass[twocolumn,twocolappendix]{aastex631} 
\usepackage{multirow}
\usepackage{amsmath}
\usepackage{tabularx}
\usepackage{booktabs}

\newcommand{\bq}{\boldsymbol q}

\newcommand{\bx}{\mathbf{x}}
\newcommand{\bk}{\textbf{k}}

\newcommand{\ihmpc}{\ h{\rm Mpc}^{-1}}
\newcommand{\mpch}{\,h^{-1}{\rm Mpc}}

\newcommand{\nbar}{\bar{n}}
\newcommand{\perr}{P_{\rm err}(k)}
\newcommand{\B}{\rule[0ex]{0pt}{0pt}}
\newcommand{\T}{\rule{0pt}{0ex}}

\begin{document}
\shorttitle{The impact of galaxy physics on field-level bias}
\shortauthors{Shiferaw, Kokron \& Wechsler}
\title{
How do uncertainties in galaxy formation physics impact field-level galaxy bias?}

\author{Mahlet Shiferaw}
\email{mahlet@stanford.edu}
\affiliation{Department of Physics, Stanford University, 382 Via Pueblo Mall, Stanford, CA 94305, USA}
\affiliation{Kavli Institute for Particle Astrophysics, Stanford University, 452 Lomita Mall, Stanford, CA 94305, USA}
\affiliation{SLAC National Accelerator Laboratory, 2575 Sand Hill Road, Menlo Park, CA 94025, USA}
\author{Nickolas Kokron}
\affiliation{School of Natural Sciences, Institute for Advanced Study, 1 Einstein Drive, Princeton, NJ, 08540, USA}
\affiliation{Department of Astrophysical Sciences, Princeton University, 4 Ivy Lane, Princeton, NJ, 08544, USA}
\affiliation{Department of Physics, Stanford University, 382 Via Pueblo Mall, Stanford, CA 94305, USA}
\affiliation{Kavli Institute for Particle Astrophysics, Stanford University, 452 Lomita Mall, Stanford, CA 94305, USA}
\author{Risa H. Wechsler}
\affiliation{Department of Physics, Stanford University, 382 Via Pueblo Mall, Stanford, CA 94305, USA}
\affiliation{Kavli Institute for Particle Astrophysics, Stanford University, 452 Lomita Mall, Stanford, CA 94305, USA}
\affiliation{SLAC National Accelerator Laboratory, 2575 Sand Hill Road, Menlo Park, CA 94025, USA}

\begin{abstract}
Our ability to extract cosmological information from galaxy surveys is limited by uncertainties in the galaxy–dark matter halo relationship for a given galaxy population, which are governed by the intricacies of galaxy formation. To quantify these uncertainties, we examine quenched and star-forming galaxies using two distinct approaches to modeling galaxy formation: \textsc{UniverseMachine}, an empirical semi-analytic model, and the \textsc{IllustrisTNG} hydrodynamical simulation. We apply a second-order hybrid N-body perturbative bias expansion to each galaxy sample, enabling direct comparison of modeling approaches and revealing how uncertainties in the galaxy–halo connection affect bias parameters and non-Poisson noise across number density and redshift. Notably, we find that quenched and star-forming galaxies occupy distinct parts of bias parameter space, and that the scatter induced from these different galaxy formation models is small when conditioned on similar selections of galaxies. We also detect a signature of assembly bias in our samples; this leads to small but significant deviations from analytic bias predictions, while assembly-bias-removed samples match these predictions well. This work indicates that galaxy samples from a spectrum of reasonable, physically motivated models for galaxy formation give a relatively small range of field-level galaxy bias parameters. We estimate a set of priors from these models that should be useful in extracting cosmological constraints from LRG- and ELG-like samples. Looking forward, careful estimates of the range of impacts of galaxy formation, for a given sample and cosmological analysis, will be an essential ingredient for extracting the most precise cosmological information from current and future large galaxy surveys.
\end{abstract}


\section{Introduction} \label{sec:intro}
Precision cosmology has entered a new era defined by an extraordinary wealth of data. Within the next decade, a multitude of ambitious new imaging surveys will see first light and map billions of galaxies. These include the Vera C. Rubin Observatory’s Legacy Survey of Space and Time \citep{2018arXiv180901669T, Ivezic_2019}, the Nancy Grace Roman Space Telescope \citep{2019arXiv190205569A, 2021MNRAS.507.1746E}, and the Spectro-Photometer for the History of the Universe, Epoch of Reionization and Ices Explorer \citep[SPHEREx]{2014arXiv1412.4872D}. Currently ongoing spectroscopic galaxy surveys, including the Dark Energy Spectroscopic Instrument \citep[DESI]{2016arXiv161100036D} and Euclid \citep{2011arXiv1110.3193L}, are beginning to provide the first ``Stage IV" cosmology constraints~\citep{desicollaboration2024desi2024vicosmological, desicollaboration2024desi2024vfullshape}. 
We can soon expect an explosion of rich, high signal-to-noise measurements of cosmological statistics at large and small scales due to the simultaneous increase in volume and density of galaxies cataloged by these surveys.
Unfortunately, modeling uncertainties of galaxy and matter density fields in the non-linear regime limit our ability to fully extract data from these new measurements
\citep[e.g.,][]{2017arXiv170609359K, 2020MNRAS.491.5498M, 2021PhRvL.126b1301P}. 

One challenge of using these new survey data across their widest range of scales is improving the robustness of our theoretical models of galaxy clustering. Improving these models requires advancing our understanding of the galaxy--halo connection, which describes the statistical and physical relationship between luminous galaxies and the dark matter halos in which they are hosted. This is the key to deciphering the physics of galaxy formation, as there is currently a wide spectrum of acceptable modeling approaches to the galaxy--halo connection, ranging from ab initio models (e.g., hydrodynamical simulations) to more computationally tractable phenomenological models (e.g., halo occupation models). See \cite{doi:10.1146/annurev-astro-081817-051756} for a broad overview of different approaches, and \cite{doi:10.1146/annurev-astro-082812-140951} for a detailed review of ab initio models only. In this study, we present a way to quantify this uncertainty in the galaxy--halo connection using a methodology
that employs the bias expansion to compare distinct galaxy formation models, as well as different populations of galaxies. Using this technique, we can place informative priors on the bias parameters \citep{2020JCAP...12..013B, 2024PhRvD.110f3538I, 2024arXiv241008998A} which can both improve constraints on cosmological parameters~\citep{2024arXiv240912937Z, 2024arXiv240910609I}, alleviate projection effects~\citep{2023JCAP...01..028C}, and inform which regimes the different galaxy formation models make similar or distinct predictions.

In its simplest form, the bias expansion states that galaxies are observable and biased tracers of dark matter \citep{1984ApJ...284L...9K, 1996MNRAS.282..347M}, which means that there is a statistical relationship between their spatial distribution and the underlying distribution of dark matter. This relationship is known as galaxy bias (see \citealt{2018PhR...733....1D} for a review), and its exact form varies depending on the tracer population. On large scales, galaxy bias can be approximated by a single proportionality factor, the linear bias term. However, to accurately describe the way galaxies are distributed today, it is necessary to model the non-linear growth of structure. This can be achieved using the bias expansion, a perturbative parameterization of galaxy bias that encodes the response of galaxies to changes in large-scale structure, in combination with cosmological perturbation theory (PT), which models the clustering of dark matter in the Universe.
 For a comprehensive review of ``standard'' PT, and the reasons why it breaks down beyond large scales, we refer to \cite{2002PhR...367....1B}. 

 Effective Field Theory (EFT) improves upon this approach with a more rigorous separation of small and large scales \citep{2012JCAP...07..051B, 2012JHEP...09..082C}.
In doing so, it pushes the regime of accuracy of perturbation theory to smaller scales. However, by construction, EFT still breaks down at non-linear scales. To model scales beyond this, phenomenological models are required: these include the halo model, as well as full numerical solutions of the equations of structure formation, commonly in the form of N-body simulations. These simulations accurately capture the small-scale structure of the Universe, but their computational expense makes it extremely difficult to explore a wide cosmological parameter space. Modeling biased tracers of dark matter, such as galaxies, in such simulations is even more computationally expensive,
motivating a technique known as Hybrid EFT (HEFT). This is a recent class of models that combines the small-scale accuracy of N-body simulations with the flexibility and generality of perturbation theories. HEFT operates in a Lagrangian coordinate system and uses the displacements from simulations rather than solving for them perturbatively as in traditional EFT \citep{2020MNRAS.492.5754M, 2021MNRAS.505.1422K, Kokron_2022, 2023MNRAS.524.2407Z}. This is the modeling approach 
we adopt in this paper, as HEFT 
has the capability to greatly improve cosmological constraints \citep{2021JCAP...09..020H}, 
 and is starting to be actively used
 to obtain cosmological constraints from real data for the first time
\citep{2024arXiv240704795C,2024arXiv240704607S}. 
 
The advent of complex bias models has presented a new challenge: the inclusion of additional nuisance parameters from the bias expansion can lead to effects in interpreting cosmological constraints in Bayesian analyses, known as prior volume effects. These are artifacts that occur when marginalizing the posterior over a highly dimensional parameter space, such that the estimated cosmological parameters are seemingly biased relative to their ``true" value \citep{2023PhRvD.108l3514H, 2023PhRvD.107l3530S, 2023JCAP...06..005M}. This is especially a problem for EFT-based analyses, which can successfully describe mildly non-linear scales, but are governed by bias parameters that are often degenerate with cosmological parameters. Although there exist techniques that circumvent this effect \citep{2023MNRAS.526.3461D, 2024MNRAS.532..783Z, 2024arXiv240500261R}, since this issue is partially driven by too-broad priors, one can instead use our knowledge of galaxy formation to determine which values of the nuisance parameters are physically allowed, and place informed priors on them to begin with. This reduces parameter space and improves computational feasibility; as degeneracies exist between the bias and cosmological parameters, tighter priors on the former allow tighter bounds on the latter. 
 
 All of this further motivates our work: by measuring the bias parameters from different galaxy formation models, we can compare various prescriptions for the galaxy--halo connection, as well as set informative priors that can reduce the volume of parameter space in survey analyses, prevent prior volume effects, and improve constraints on cosmological parameters. 

However, measuring bias parameters for this purpose is still in its infancy and placing priors on them even more so. There is a wide variety of techniques used to fit the bias parameters (forward models, summary statistics, field-level inference, normalizing flows), the types of tracers studied (galaxies, halos), and the bias expansion model (Eulerian, Lagrangian, hybrid) that is employed. 

On the halo bias side, \cite{2016JCAP...02..018L}, \cite{2018JCAP...09..008L} and \cite{2021JCAP...10..063L} measured the bias parameters of dark matter halos from gravity-only simulations using a bias expansion in the Eulerian coordinate frame. Although these results are useful for forecasts based on halo statistics and inform analytic models of halo formation, it is necessary to study {\em galaxy} bias if the goal is to constrain parameters using galaxy clustering. To this end, \cite{2021JCAP...08..029B} and \cite{2024arXiv240213310I} set priors on Eulerian bias parameters from galaxies in a hydrodynamical simulation and halo occupation distribution (HOD) model, respectively. \cite{2022MNRAS.514.5443Z} placed priors using a hybrid Lagrangian bias expansion like ours, using a model based on extended subhalo abundance matching. 
Our study extends and complements these previous works by exploring the impact of galaxy formation model, galaxy type, and redshift on the bias parameters, particularly the relation between non-linear and linear bias. Shortly before our submission, \cite{ivanov2024millenniumastridgalaxieseffective} posted a similarly scoped study to the arXiv; we comment further on this work in \S\ref{sec:note}.

In this work, we compare different models of galaxy formation using HEFT to second order in the Lagrangian bias expansion, plus a cubic term. Building upon \cite{Kokron_2022}'s technique for measuring and setting priors on stochasticity from HEFT, we avoid the need to sample a large parameter space by employing an inverse-modeling approach using maximum likelihood. With this method, we fit the bias parameters to galaxy samples from two distinct models:  \textsc{IllustrisTNG} \citep{nelson2021illustristng} and \textsc{UniverseMachine} \citep{2019MNRAS.488.3143B}, as well as assembly-bias-removed versions of each. Together, these can be considered representative of the range of empirical and physical approaches that can differently impact the galaxy--halo connection and galaxy bias.
In each model, we create samples of quenched and star-forming galaxies, motivated by DESI-like Luminous Red Galaxies (LRGs) and Emission Line Galaxies (ELGs), respectively. Our aim is not only to understand the clustering of these different galaxy populations relative to the clustering of dark matter, as parameterized via the bias expansion, but also to understand the variation in bias parameters that results from changes in sample, redshift, and galaxy formation model. We additionally provide reasonable and physically informed priors on their bias parameters for future use in EFT-based cosmological constraint analyses. 

The remainder of this paper is structured as follows. 
In \S\ref{sec:theory}, we describe the HEFT model and our technique to fit the bias parameters. In \S\ref{sec:models}, we summarize the galaxy formation models and outline our methodology for creating the galaxy samples used in this work. We present the measured values of the bias parameters in \S\ref{sec:results}, and analyze our results in \S\ref{sec:discussion} by making a comparison to halo-model-based analytic predictions. We additionally quantify the impact of assembly bias and examine the differences between each galaxy model and each galaxy sample. We discuss the priors we set on the bias parameters, examine the degree of non-Poisson stochasticity in our samples, and study the redshift dependence of linear bias for these samples. We summarize our findings, as well as lay out future directions in  \S\ref{sec:conclusions}. 

\section{Theory}
\label{sec:theory}

We employ the techniques utilized in \cite{Kokron_2022} to study the bias parameters for a varying class of tracer--matter connection models. A short overview of the bias expansion we employ is presented in \S\ref{sec:HEFT} and \S\ref{sec:measurement}.

\subsection{The bias expansion}
\label{sec:HEFT}

In its most general form, galaxy bias can be written as the following:

\begin{equation}
\label{eq:bias}
    \delta_g(\textbf{x},\tau)=\sum_i b_i(\tau) \mathcal{O}_i(\bx,\tau),
\end{equation}
where $\delta_g(\mathbf{x}) \equiv \frac{n_g(\mathbf{x})}{\bar{n}_g}-1$ is the galaxy density contrast, $b_i(\tau)$ is the bias parameter at the $i$th order, $n_g$ is the number density of galaxies, and $\tau$ is some suitable definition of time (e.g., scale factor $a$, redshift $z$, conformal time $\eta$, or coordinate time $t$). 
Equation \ref{eq:bias} thus relates the galaxy density contrast $\delta_g(\mathbf{x},\tau)$ to a sum of operators $\mathcal{O}_i(\textbf{x}, \tau)$, akin to a set of basis fields that are 
composed of different scalar combinations of the Hessian matrix of the gravitational potential $\partial_i \partial_j \Phi(\bx,\tau)$. As a result, each operator $\mathcal{O}_i(\bx , \tau)$ is either a function of the matter density $\delta_m(\mathbf{x}, \tau)$ or the tidal field strength $s_{ij} = (\partial^{-2} \partial_i \partial_j - \delta_{ij}/3) \delta_m$ and derivatives thereof\footnote{There are also non-local in time contributions that can arise, but at higher order than considered here~\citep{Senatore_2015}.}. These statistical fields describe a certain property of the underlying dark matter distribution that the galaxy density contrast depends on and are accordingly weighted by an associated bias parameter $b_i(\tau)$. 

Under this generalized bias expansion, we can choose the Eulerian or Lagrangian coordinate frame to work within. The Eulerian bias expansion connects the \emph{current-time} basis fields to the current-time galaxy density contrast. The Lagrangian bias expansion, on the other hand, establishes the relation between tracers and matter in the initial conditions of the Universe and assumes it is preserved, evolving only due to the underlying dynamics of the matter fluid. In this paper, we use the Lagrangian picture, which entails remapping each final Eulerian coordinate $\mathbf{x}$ in terms of its initial Lagrangian position $\mathbf{q}$ and a Lagrangian displacement vector $\mathbf{\Psi}(\mathbf{q}, \tau)$:

\begin{equation}
\label{eq:lagrangian}
    \mathbf{x}(\mathbf{q}, \tau)=\mathbf{q}+\mathbf{\Psi}(\mathbf{q, \tau}).
\end{equation}

The relationship between the Lagrangian galaxy density contrast $\delta_g$ and the Hessian of the potential at the initial conditions is written as a combination of the functional $F$ and a stochastic contribution $\epsilon$:
\begin{alignat}{2}
    \delta_g (\bq) &= &&F[\partial_i \partial_j \Phi (\bq)] + \epsilon (\bq) \\
    &\approx &&1 + b_1 \delta_L + b_2 \left ( \delta_L^2 - \sigma_L^2 \right ) + \\
    &\nonumber  &&b_s^2 \left ( s^2_L - \frac{2}{3} \sigma_L^2 \right) + b_{\nabla^2} \nabla^2 \delta_L + \epsilon,
\end{alignat}
where in the second line, we have expanded this functional to second order in density fluctuations, as has been commonly adopted in Lagrangian bias studies~\citep{Vlah_2016}. Through this time evolution process, the proto-galaxy density fluid elements at early times are advected to their late-time positions. Their density contrast is therefore written as 
\begin{align}
\label{eqn:advectbias}
    1+\delta_g(\mathbf{x}, \tau)=\int d^3 q\, &F[\partial_i \partial_j \Phi (\bq, \tau)] \times \\
    \nonumber &\delta^{D}\left (\mathbf{x}(\mathbf{q}, \tau)-\mathbf{q}-\mathbf{\Psi}(\mathbf{q}, \tau)\right ),
\end{align}
where the Lagrangian displacement vector $\mathbf{\Psi}(\mathbf{q}, \tau))$ is then expanded perturbatively around small displacements such that $\mathbf{\Psi}=\mathbf{\Psi}^{(1)}+\mathbf{\Psi}^{(2)}+\mathbf{\Psi}^{(3)}+\cdots$ \citep{PhysRevD.77.063530}. LPT thus perturbatively determines the properties of $\mathbf{\Psi}(\mathbf{q}, \tau)$, as well as summary statistics for $\delta_g(\textbf{x}, \tau)$. 

\cite{2020MNRAS.492.5754M} point out that N-body dark matter simulations also solve for the displacement vector $\mathbf{\Psi}(\mathbf{q})$, albeit non-perturbatively. Taking advantage of this fact naturally leads to
Hybrid Effective Field Theory (HEFT), a field-level model that combines the analytic bias expansion of Equation \ref{eq:bias} with numerical displacements from an N-body simulation in order to obtain the late-time tracer field \citep{2021MNRAS.505.1422K, 2021arXiv210112187Z, 2021JCAP...09..020H}.
In the context of HEFT, the operators $\mathcal{O}_i(\textbf{x}, \tau)$ in Equation \ref{eq:bias} are constructed at late times via advection, using the displacement vector $\mathbf{\Psi}(\mathbf{q}, \tau)$ from N-body simulations.

In this work, we represent the hybrid bias expansion for the galaxy tracer field at late times as

\begin{equation}
\label{eq:bias expansion}
\begin{split}
\delta_g(\mathbf{x})= & \delta_m(\mathbf{x})+b_1 \mathcal{O}_\delta(\mathbf{x})+b_{\nabla^2} \mathcal{O}_{\nabla^2 \delta}(\mathbf{x})+b_2 \mathcal{O}_{\delta^2}(\mathbf{x})+ \\
& b_{s^2} \mathcal{O}_{s^2}(\mathbf{x})+ b_3\mathcal{O}_{\delta^3}(\mathbf{x})+\epsilon(\mathbf{x}),
\end{split}
\end{equation}
where a given operator $\mathcal{O}_i(\mathbf{x})$ corresponds to the result of advecting 
$F_i(\mathbf{q})$ ($F_i \supset 1, \delta_L, \delta_L^2, s_L^2, \nabla^2 \delta_L,...$)
through the integral in Equation~\ref{eqn:advectbias}. Re-expressing Equation~\ref{eq:bias expansion} in terms of Eulerian bias fields, using Equation~\ref{eqn:advectbias} and LPT expressions for displacements, leads to the co-evolution relations between Eulerian and Lagrangian bias~\citep{2016JCAP...02..018L, Abidi_2018}. Alternatively, exponentiating the linear displacements and expanding higher-order displacements leads to the definition of the \emph{shifted operator basis} of \cite{Schmittfull_2019}. The component fields $\mathcal{O}_i(\textbf{x}, \tau)$ in this hybrid EFT expansion are visualized at $z=0.5$ in Figure \ref{fig:expansion}.

We have also included a stochastic field $\epsilon(\mathbf{x}, \tau)$ with vanishing mean. This $\epsilon$ field accounts for the fact that the bias expansion is both truncated at some order and not fully deterministic. Very small-scale physics, i.e., on scales below $k_h \sim R_L^{-1} \sim 0.4 \mpch$ ($R_L$ is the Lagrangian radius of a halo of mass M) where this bias expansion is applicable, introduces scatter in the relation. The $\epsilon$ field itself admits a perturbative 
description and may also be expanded in a basis of operators that are explicitly uncorrelated with the deterministic fields 
$\mathcal{O}_i$~\citep{2018PhR...733....1D}. We refer the reader to \cite{Kokron_2022} for further discussion of the two-point statistics of this stochastic field in the context of the HEFT expansion. 

 In addition to the standard second-order Lagrangian basis of bias fields considered in past work, we have also added a local cubic operator $\delta^3$ to include contributions to the stochasticity from third-order operators. While there are nominally four cubic operators --- $\delta^3, \delta s^2,$ and $s^3$, as well as a ``non-local" operator $s_{ij}(\bq) t^{ij}(\bq)$\citep{2018JCAP...09..008L} --- their contributions are degenerate at the level of one-loop perturbation theory for the power spectrum. Including the cubic operator $\delta^3$ thus helps to capture excess super-Poisson stochasticity that could improve our fit of the other bias parameters. However, we note that by \emph{only} including $\delta^3$ and not the other third-order parameters, the associated bias does not necessarily correspond to the traditional peak-background split bias of the local $\delta^3$ field. Instead, it may also possess contributions from other cubic operators degenerate with $\delta^3$. 
 
 When realizing the $\delta^3$ field in the Lagrangian expansion, we subtract its overlap with the standard linear density such that 
\begin{align}
    \langle \delta (\bk) \delta^3 (\bk') \rangle' = 3 \sigma^2 P(k), \quad \delta^3 \to \delta^3(\bq) - \sigma^2 \delta^3 (\bq),
\end{align}
where $\sigma^2 = \langle \delta^2 \rangle$ is the variance of the initial conditions field. This redefinition removes ambiguity between the value of $b_1$ and $b_3$ at low $k$. 
Due to past numerical challenges in realizing the $\nabla^2 \delta$ field in the context of HEFT
\citep{2021MNRAS.505.1422K, 2021JCAP...09..020H},
we also apply a small-scale damping of power by an exponential function $e^{-(kR)^2/2}$. Since we are only interested in mitigating numerical issues, we choose $R=1 \mpch$ as the Lagrangian radius over which to smooth the $\nabla^2 \delta_L$ field; this only affects the field on scales smaller than those where the bias expansion is expected to hold.

\subsection{Measuring the bias parameters}
\label{sec:measurement}

Next, we outline our procedure for measuring the bias parameters of a galaxy tracer field. We adopt the approach detailed in \cite{Kokron_2022}; see \S2.3 of that work for a thorough description of the methodology. In summary, we transform the galaxy tracer field (defined in \S\ref{sec:HEFT}) into Fourier space:
\begin{equation}
    \delta_g(\mathbf{k})=\delta_m(\mathbf{k})+\sum_i b_i \mathcal{O}_i(\mathbf{k})+\epsilon(\mathbf{k}),
\end{equation}
and solve for the stochasticity field $\epsilon(\mathbf{k})$ as

\begin{equation}
\label{eq:stochastic}
\epsilon(\mathbf{k})=\delta_g(\mathbf{k})-\delta_m(\mathbf{k})-\sum_i b_i \mathcal{O}_i(\mathbf{k}).
\end{equation}

This stochasticity field is characterized by a vanishing mean and is real-valued in configuration space such that $\epsilon^\ast(\mathbf{x})=\epsilon(\mathbf{x})$ and $\epsilon^\ast(\mathbf{k})=\epsilon(-\mathbf{k})$. We may describe the two-point structure of the stochasticity by studying its power spectrum:
\begin{equation}
\label{eq:Perr}
    \langle\epsilon(\mathbf{k}) \epsilon(\mathbf{k}')\rangle \equiv (2\pi)^3 \delta^D (\bk + \bk') P_{\mathrm{err}}(k) .
\end{equation}

Solving for the bias parameters that, at a given $k$, minimize this stochasticity leads to the definition of the \emph{bias transfer functions}, $\hat{\beta}_i(k)$, defined in \cite{Schmittfull_2019}. The $k$-evolution of the bias parameter in \cite{Schmittfull_2019} is ascribed to, for example, higher-order displacement contributions that are degenerate with the linear density, and the authors compute them perturbatively in some cases. \par 
Because in HEFT, we are not interested in said contributions, and we take the bias parameters to be \emph{truly} constant in scale\footnote{The values of biases are expected to depend on the initial cut-off scale used to forward-model galaxy samples~\cite[for some discussion on this]{rubira2023galaxybiasrenormalizationgroup}. We match our Lagrangian fields to the same grid scale used to initialize power in the initial conditions and treat them as constants, defined at this scale, as a result. }, we must instead define an estimator for the bias parameters that weighs large and small scales equally to result in a single value. In this case, a change in the measurement of $\hat{b}_i$ as a function of scale is interpreted as either a breakdown of the specific bias parameterization adopted or of the bias expansion as a whole. 

We define a stochasticity field that is filtered with a Fourier-space top-hat filter, such that the field is zero for modes with $ | \mathbf{k} | > {k_\mathrm{max}}$. We refer to this smoothed field as $[\epsilon(\mathbf{x})]_{k_\mathrm{max}}$, where the subscript implies that integrals over Fourier space are truncated beyond $k_\mathrm{max}$.
We then solve for the bias parameters that minimize the overall variance of the configuration-space stochasticity
\begin{equation}
\label{eq:loss}
    S=\left\langle[\epsilon(\mathbf{x})]_{k_{\max }}^2\right\rangle.
\end{equation}

Minimizing the field-level variance is a linear problem equivalent to least-squares fitting. It can be solved analytically, from which we find the estimator for bias parameters
\begin{equation}
\label{eq:solution}
    \hat{b}_i(k)=M_{i j}^{-1} A_j,
\end{equation}
where 
\begin{equation}
\begin{split}
A_j & =\left\langle\left[\mathcal{O}_j(\mathbf{x})\left(\delta_g(\mathbf{x})-\delta_m(\mathbf{x})\right)\right]_{k_{\max }}\right\rangle \\
& =\frac{1}{V}\int_{|\mathbf{k}|<k_{\max }} \frac{d^3 k}{(2 \pi)^3} \mathcal{O}_j(\mathbf{k})\left[\delta_g-\delta_m\right]^*(\mathbf{k})
\end{split}
\end{equation}
and
\begin{equation}
\label{eq:M_ij}
\begin{split}
M_{i j} & =\left\langle\left[\mathcal{O}_i(\mathbf{x}) \mathcal{O}_j(\mathbf{x})\right]_{k_{\max }}\right\rangle \\
& =\frac{1}{V}\int_{|\mathbf{k}|<k_{\max }} \frac{d^3 k}{(2 \pi)^3} \mathcal{O}_i(\mathbf{k}) \mathcal{O}_j^*(\mathbf{k}) .
\end{split}
\end{equation}
The resulting estimates of the bias parameters $\hat{b}_i(k)$ include information up until some maximum $k_\mathrm{max}$ scale. The covariance of this estimator is consequently 
\begin{equation}
\label{eq:cov}
\begin{split}
    {\rm Cov}(\hat{b}_i, \hat{b}_j) & = \langle \hat{b}_i \hat{b}_j \rangle - \langle \hat{b}_i\rangle \langle \hat{b}_j\rangle\\
    & \approx \frac{1}{\bar{n} V} \left [ \boldsymbol{M}(k_{\rm max})^{-1} \right ]_{ij}.
\end{split}
\end{equation}
The derivation of this result can be found in Appendix C of \cite{Kokron_2022}\footnote{We note that the original formulation of the loss function in ~\cite{Kokron_2022} contained a missing factor of $\frac{1}{V}$. This does not affect the estimator but does affect its covariance, and we have corrected for this typo in Equation~\ref{eq:cov}.}. 

\begin{figure*}
    \centering
    \includegraphics[width=\textwidth]{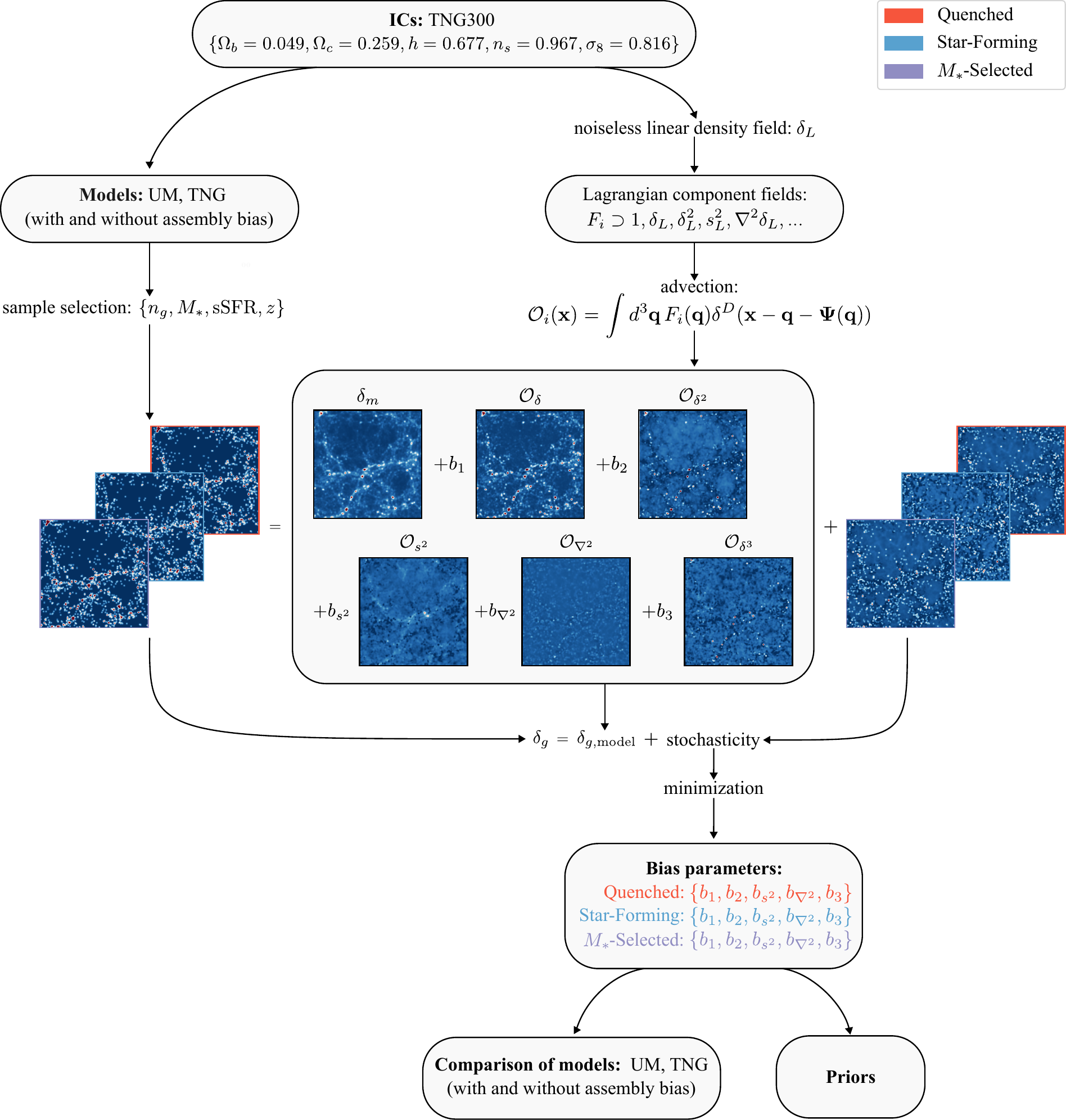}
    \caption{Visualization of the methodology of this work: starting from the same initial conditions (TNG300), we obtain both the galaxy density field $\delta_g$ for each model (UM and TNG, with and without assembly bias)
    and the noiseless linear density field $\delta_L$. Using the latter, we construct the bias-weighted fields $\mathcal{O}_i$ at redshift $z$ via advection. These constitute a deterministic model $\delta_{g,\mathrm{ model}}$ that differs from the galaxy density field $\delta_g$ only
    by a stochastic term $\epsilon$ (see Equation \ref{eq:bias expansion}), which we minimize to solve for the best-fit bias parameters of each sample. Finally, we compare results between the models and set priors for use in cosmological inferences. In this figure, we visualize the density field $\delta_g$ and stochasticity $\epsilon$ at $z=0.5$ for a high number density sample in TNG.}
    \label{fig:expansion}
\end{figure*}

In the remainder of this text, we assume that all of our bias parameter measurements result from this estimator $\hat{b}_i$. To improve readability, we henceforth refer to the estimator for bias parameters as simply $b_i$. Our methodology for obtaining this estimate, as well as our subsequent analysis, is summarized in Figure \ref{fig:expansion}.

\section{Simulations}
\label{sec:models}

Using the techniques outlined in \S\ref{sec:theory}, we can now compare different models of galaxy formation. The models and samples we examine are detailed in \S\ref{sec:um_tng} and \S\ref{sec:samples} below.

\subsection{Models of galaxy formation}
\label{sec:um_tng}

We use two primary galaxy formation models in this study: \textsc{UniverseMachine} (UM), an empirical semi-analytic model, and \textsc{IllustrisTNG} (TNG), a hydrodynamical model.
To directly compare UM and TNG's modeling approaches, we create matched galaxy samples from each model, as explained in \S\ref{sec:samples}. We also create matched samples from an additional model of the galaxy--halo connection:
a mass-dependent, non-parametric HOD that removes galaxy assembly bias from both UM and TNG, as detailed in \S\ref{sec:assembly}.
From these sample catalogs, we obtain the galaxy overdensity field $\delta_g(\mathbf{k})$, which we use to solve for the best-fit bias parameter estimator $ \hat{b}_i(k)$, as written in Equations \ref{eq:solution}-\ref{eq:M_ij}. To make a direct comparison between the bias parameters measured from the two models, we set identical initial conditions by running the \textsc{UniverseMachine} DR1 model on the non-baryonic physics ``dark matter only" counterpart to TNG300, known as TNG300-1 Dark, with the following cosmological parameters: \{$\Omega_b=0.0486, \Omega_c=0.2589, h=0.6774, n_s=0.9667, \sigma_8=0.8159$\}. This ensures that the large-scale structure is identical in the UM and TNG boxes.

The \textsc{UniverseMachine} (UM) is an empirical model of galaxy formation that links star formation to properties of the host halo \citep{2019MNRAS.488.3143B}. Specifically, UM models galaxy formation and evolution by constraining the star formation rate (SFR) of a galaxy as a function of the dark matter halo’s maximum circular velocity, redshift, and accretion history. UM thus assigns an SFR distribution to
each galaxy from this 44-dimensional parameter space and integrates the SFR along the merger tree of each galaxy's host halo. From this, UM obtains a stellar mass and UV luminosity and compares the statistics of these observables to that of the real universe to obtain a likelihood of the original guess. This likelihood is fed to a Markov Chain Monte Carlo algorithm, which returns the posterior distribution of the model parameters, and the resulting best-fitting parameters for the SFR distribution.

In comparison, the \textsc{IllustrisTNG} project (TNG) is a full gravo-magnetohydrodynamic cosmological simulation that explicitly includes prescriptions for gas cooling, star formation, stellar feedback,
and supermassive black hole growth and feedback. TNG uses the moving-mesh hydrodynamic and gravity N-body code \textsc{arepo} \citep{2010MNRAS.401..791S} to model galaxy evolution and obtains initial conditions at $z=127$ using the \textsc{N-GenIC} code \citep{2005Natur.435..629S}. Dark matter halos are identified as groups using the friends-of-friends (FoF) algorithm \citep{1985ApJ...292..371D}. Galaxies are consequently defined as subhaloes within each FoF group via the \textsc{subfind} algorithm \citep{2001MNRAS.328..726S, 2009MNRAS.399..497D}. We consider the largest run in their simulation suite, TNG300, part of their publicly available data release \citep{nelson2021illustristng}. This box has a size of $L_\mathrm{box}=205 \mpch$, which is $\sim300$ Mpc, with 2500$^3$ dark matter and gas particles each.

UM and TNG adopt fundamentally different approaches to modeling galaxy formation. UM employs an iterative MCMC framework to calibrate against a wide array of observational data—including galaxy stellar mass functions (SMFs), specific star formation rates (sSFRs), quenched fractions, correlation functions, and more—producing a realistic distribution of galaxy star formation rates (SFRs). By relaxing theoretical priors in favor of empirical constraints, UM is best described as an empirical model. In contrast, TNG is an \textit{ab initio} simulation that evolves the matter distribution using hydrodynamical equations while modeling complex astrophysical processes through sub-grid prescriptions. Although TNG is tuned to match several key observables, such as the galaxy SMF, the stellar-to-halo mass relation, and the cosmic SFR density, its sub-grid models are informed by data but are not rigorously marginalized over, and thus reflect strong physical priors. We touch upon the potential impact of these underlying differences between UM and TNG on the measured bias parameters in \S\ref{sec:galaxy model comparison}.

\subsection{Selecting galaxies at a target number density}
\label{sec:samples}

We create matched samples of galaxies from each galaxy formation model (UM and TNG) to mimic the target number density and redshift spanned by DESI's selection of Luminous Red Galaxies (LRGs) and Emission Line Galaxies (ELGs) \citep{2016arXiv161100036D}.
LRGs are early-type, elliptical galaxies with red rest-frame optical wavelengths that occupy the redshift range $z\approx[0.4,1.0]$. ELGs are late-type, spiral galaxies with blue rest-frame optical wavelengths that occupy the redshift range $z\approx[0.6,1.6]$. 
We choose to match DESI's LRG target number density of $\bar{n}_g \approx 5\times10^{-4}$ $[\mpch]^{-3}$ \citep{2023AJ....165...58Z, 2022MNRAS.512.5793Y}, as it is larger than the target ELG number
density in the same redshift range \citep{2023AJ....165..126R}. 

To encompass a significant range of the redshifts for DESI's LRG and ELG target galaxies, our data comprises snapshots at redshifts $z=\{0.0,0.5,1.0,1.5\}$.
At each of these snapshots, 
we adopt a threshold-based cut in stellar mass ($M_*$) and
sSFR. We emphasize that these cuts act as a proxy for galaxy color, so that we can select samples in our study that are similar to DESI's LRGs and ELGs without explicitly making cuts in color--magnitude space. We thereby create populations that mimic quenched galaxies with little-to-no active star formation, as well as galaxies with currently ongoing star formation \citep{2009ApJ...707..250M}, in rough correspondence to the LRGs and ELGs that DESI targets, respectively. For completeness, we also use the $M_*$ cuts to create a general $M_*$-selected population; this combined sample of the quenched and star-forming galaxies is treated as an additional galaxy type in our study.

\begin{figure*}
    \centering
    \includegraphics[width=\textwidth]{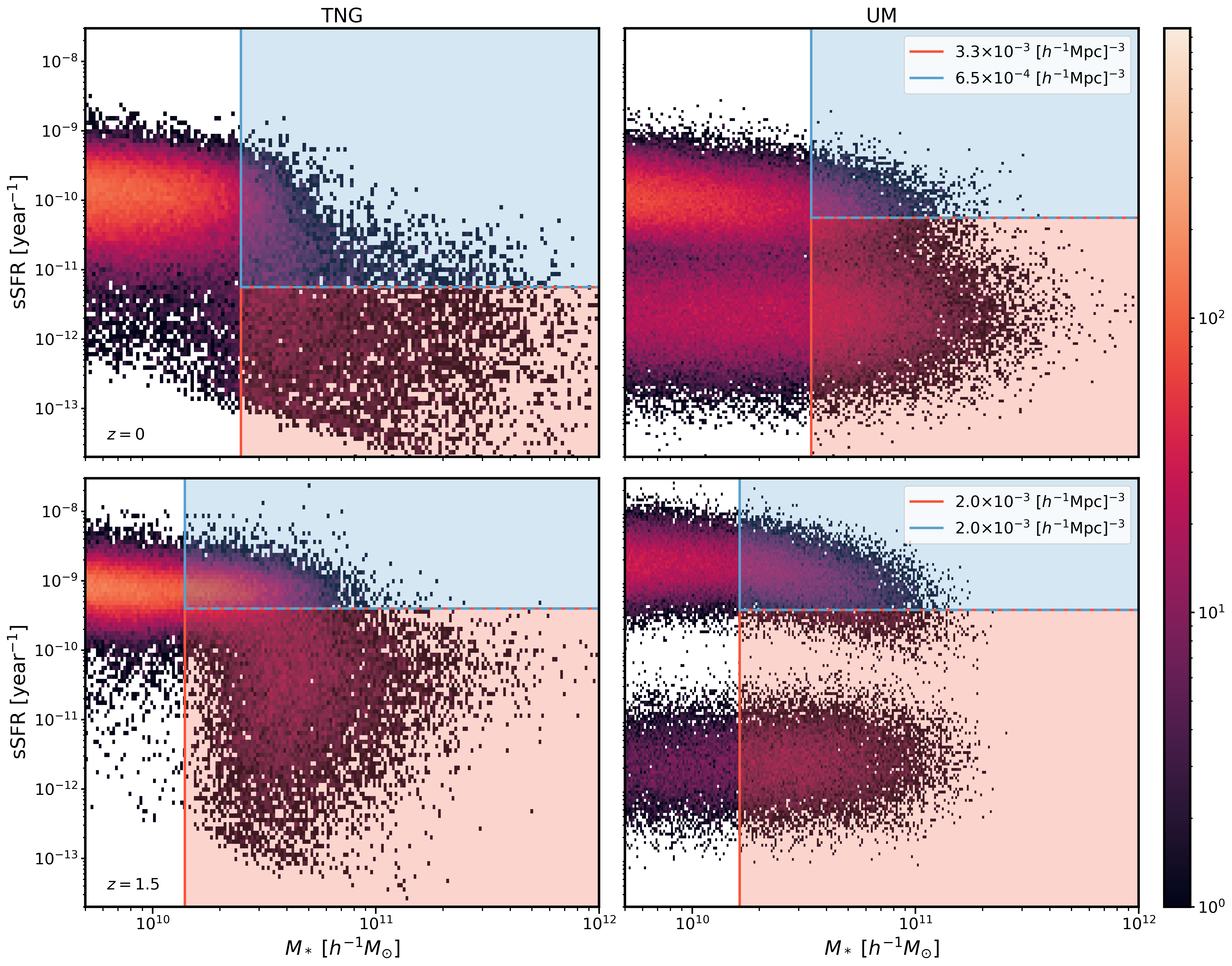}
    \caption{A 2D histogram of $M_*$ and sSFR for TNG and UM galaxies at redshifts $z=0$ (top panel) and $z=1.5$ (bottom panel). The red and blue shaded regions mark the $M_*$ and sSFR cuts which separate the quenched and star-forming samples, respectively. The number density for each population is shown in the legend (medium number density, medium $M_*$).}
    \label{fig:sample}
\end{figure*}

To account for variation in the number density achieved by future surveys, we choose three target number densities for each galaxy formation model at all redshifts. These include a ``low" number density ($10^{-3}$ $ [\mpch]^{-3}$), corresponding to a realistic, conservative cut that matches the number density of our quenched and star-forming galaxies to DESI's target number density for LRGs, as well as a medium (3.9$\times10^{-3}$ $ [\mpch]^{-3}$) and high number density cut (6.8$\times10^{-3}$ $ [\mpch]^{-3}$), which are increasingly more ambitious to achieve in very wide-area galaxy redshift surveys.
Overall, we investigate 144 galaxy samples: three selections in number density; quenched, star-forming, and $M_*$-selected galaxies; UM and TNG, as well as assembly-bias-removed versions of each; and four redshift bins.
The number densities of each sample, as well as the thresholds in $M_*$ and sSFR that define each galaxy type, are detailed in Table \ref{tab:samples} of the Appendix. A visual illustration of our selection process, which we describe below, as well as the distribution of sSFR and $M_*$ in the TNG and UM galaxy formation models, is shown in Fig.~\ref{fig:sample}.

To create these samples, we first reach our chosen target number density by making a cut in $M_*$. As listed in the ``$M_*$-selected" column of Table \ref{tab:samples}, these target number densities consist of a low, medium, and high-density cut, which entails a high, medium, and low stellar mass cut respectively. Then, to distinguish between star-forming and quenched galaxies at redshifts $z=\{0.5,1.0,1.5\}$, and due to the tight relationship between star formation rate (SFR) and $M_*$, we divide the $M_*$-selected sample of galaxies in half by making a cut in sSFR; this is a common technique in observations and simulations alike for separating quenched and star-forming galaxies, and is typically made at sSFR=$10^{-11}$ year$^{-1}$ \citep{2010ApJ...714L..79C, 2021MNRAS.500.2036K, 2025arXiv250116545S}. This cut in sSFR, which we calculate using each galaxy simulation's instantaneous SFR and $M_*$, is depicted by the horizontal dashed blue and red lines in Figure \ref{fig:sample}. We thus ensure that each population has identical number densities, which is optimal for a direct comparison because the covariance of the bias parameter measurements is inversely proportional to the number density of the sample (see Equation \ref{eq:cov}). We repeat this process for all our target number densities.

The samples at redshift $z=0$ follow a slightly modified procedure, as there are not enough star-forming galaxies in TNG to evenly split the samples at this redshift. 
We instead divide the $M_*$-selected sample of galaxies so that there are five times as many quenched galaxies as star-forming galaxies ($n_{g\mathrm{, quenched}} = 5 n_{g\mathrm{, star-forming}})$, such that the number density of star-forming galaxies are closer to our density selections at other redshifts. This is reflected in the target number densities listed for each sample in the ``Quenched" and ``Star-Forming" columns of Table \ref{tab:samples}. 
It is important to note that while we achieved an even split of quenched and star-forming galaxies in UM, we chose to repeat this modified procedure for both simulations. This is so that we could maintain identical conditions for both models and ultimately enable a direct comparison of our bias measurements. 

Figure \ref{fig:sample} depicts our selection of quenched and star-forming galaxy samples in the $M_*$-sSFR parameter space of UM and TNG with red and blue shaded regions. To show how the galaxy catalog evolves with redshift in both simulations, the first row shows our selections at $z=0$, and the second row shows our selections at $z=1.5$. As the red-shaded region in the upper left panel visualizes, the sSFR cut for quenched galaxies in TNG at $z=0$ is not unreasonably high,
showing that this modified selection procedure is a suitable one. Similarly, the upper right panel makes it clear that there are more quenched galaxies than star-forming galaxies in UM at $z=0$ due to this modified selection procedure, as indicated by the higher sSFR cut.

We note that there are many galaxies with zero SFR in TNG, particularly at redshift $z=0$. This likely contributes to the overabundance of quenched galaxies that we find in TNG at $z=0$, as roughly $67-70\%$ of the galaxies in our selection at this snapshot are assigned a SFR of zero, compared to $\sim6\%$ at $z=1.5$, and $<0.5\%$ of the UM galaxy samples across all redshifts and number density cuts.
To avoid numerical errors from taking the log of zero SFR, we offset the SFR of each galaxy by a trace amount ($10^{-13}\, M_\odot (h{\rm yr})^{-1}$). As a result, these zero-SFR galaxies are not visible in Figure \ref{fig:sample}, but are included in the quenched galaxy sample. For consistency, we implement this offset for all samples at redshifts $z=\{0.0, 0.5, 1.0, 1.5\}$ in both TNG and UM. 

Finally, we point out that this $M_*$, sSFR threshold-based cut omits any assumption about the galaxy's color magnitude or emission line luminosity, as the selection criteria for these observables often vary between surveys. For instance, Euclid and the Nancy Grace Roman Space Telescope are targeting H$\alpha$ and [O III]-selected ELGs, while DESI targets [O II] emitters. So while our study is indeed motivated by the LRGs and ELGs targeted by DESI, our sample selection method does not impose DESI-specific cuts. As a result, the bias parameter measurements we discuss in \S\ref{sec:results}, and the corresponding priors which we present in \S\ref{sec:priors}, can safely generalize to other surveys, so long as they target $M_*$-selected galaxies, LRGs, and/or ELGs  in a similar redshift and number density range covered by our study.

\section{Results}
\label{sec:results}

Having created the galaxy samples at each redshift and target number density for both the TNG and UM galaxy formation models, we now find the best-fit bias parameters using the procedure described in \S\ref{sec:measurement}. The key measurements are presented in \S\ref{sec:measurements} in the form of the $b_2(b_1)$, $b_{s^2}(b_1)$, $b_{\nabla^2}(b_1)$, and $b_3(b_1)$ bias relations. In \S\ref{sec:galaxy model comparison}, we make a 1:1 comparison of these measured bias parameters across 
the galaxy formation models. 

\subsection{The bias relations}
\label{sec:measurements}

\begin{figure*}
    \centering
    \includegraphics[width=\textwidth]{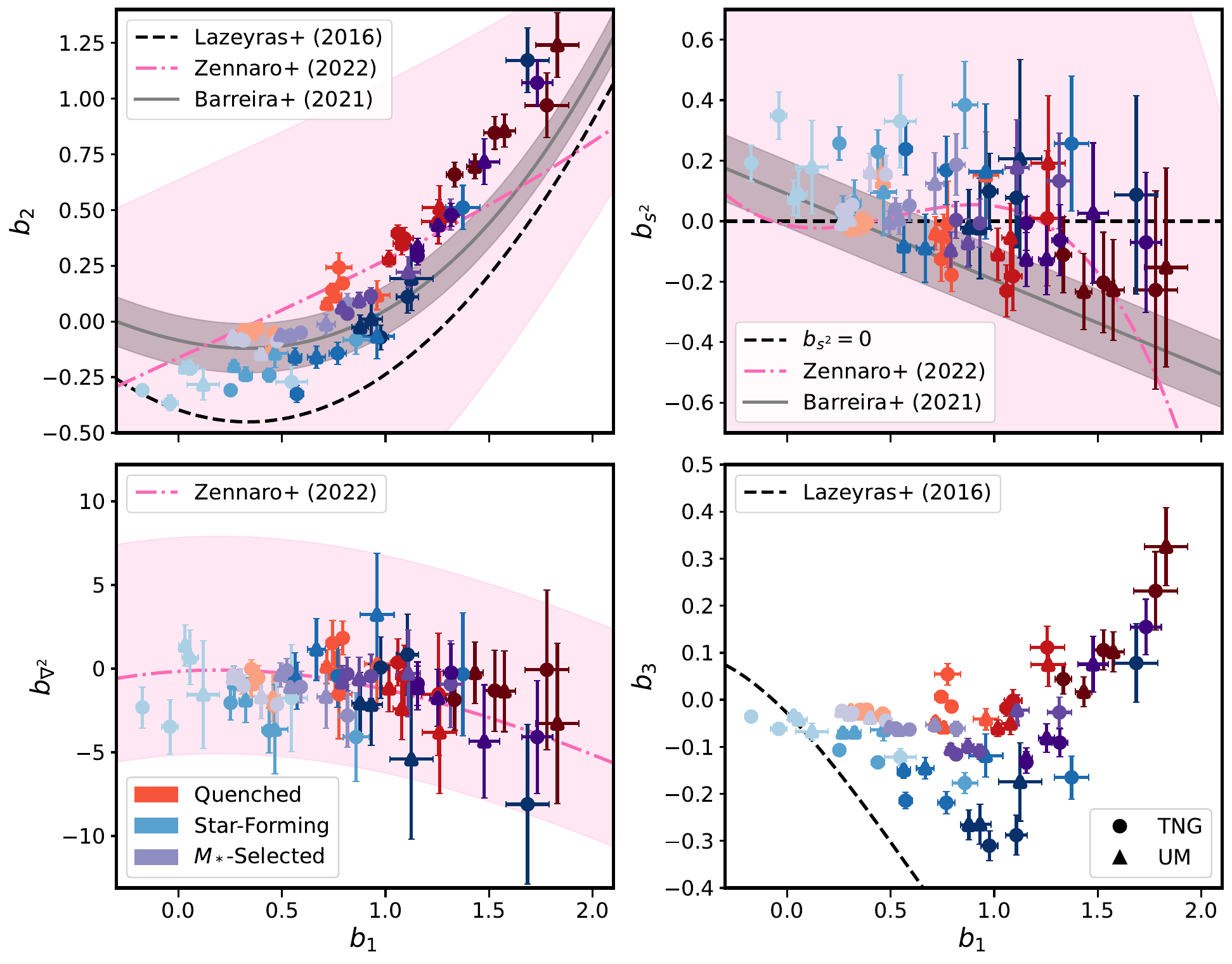}
    \caption{
    The bias parameter relations as a function of $b_1$ at $z=\{0.0, 0.5, 1.0, 1.5\}$ for each sample of galaxies in UM (triangles) and TNG (circles), including quenched (red), star-forming (blue), and the $M_*$-selected population (purple). The color intensity (light to dark) maps to the redshift (low to high). The error bars are calculated according to the steps outlined in Appendix C of \cite{Kokron_2022}. 
    Best-fit relations for halos~\citep{2016JCAP...02..018L} and galaxies~\citep{2021JCAP...08..029B, 2022MNRAS.514.5443Z} from past works are shown as comparison.  
    } 
    \label{fig:bias}
\end{figure*}

To compare our measurements to those in the literature, we examine the $b_2(b_1)$, $b_{s^2}(b_1)$, $b_{\nabla^2}(b_1)$, and $b_3(b_1)$ bias relations. Using the best-fit bias parameter solution in Equation \ref{eq:solution}, we can obtain measurements of the bias parameters for each galaxy sample at a chosen $k_\mathrm{max}$. We use $k_\mathrm{max}=0.2 \ihmpc$ as conservative estimate to report our final results, as it has been shown that perturbative approaches like our own are successful in describing the clustering of galaxies down to these scales \citep{2016JCAP...03..017B}. In Appendix \ref{app:consistency} we check that
we recover the expectation of $b_1$ from the cross-spectrum definition of the linear bias. We additionally show, in Appendix~\ref{app:consistency}, that until $k_{\rm max} = 0.2 \ihmpc$, we do not find a strong running with scale for our measured parameters.
Once we calculate the bias parameters at $k_\mathrm{max}=0.2 \ihmpc$ for each of our galaxy samples, we plot the higher-order bias parameters as a function of the linear bias parameter $b_1$. We refer to these as the bias relations.

These measurements are presented in Figure \ref{fig:bias} for UM and TNG as triangles and circles, respectively. In each panel, we plot the measured bias relations $b_2(b_1)$, $b_{s^2}(b_1)$, $b_{\nabla^2}(b_1)$, and $b_3(b_1)$ for quenched (red), star-forming (blue), and the $M_*$-selected galaxy population (purple). Each data point represents a sample of galaxies selected from redshifts $z=\{0.0, 0.5, 1.0, 1.5\}$ using the target number density cut method described in \S\ref{sec:samples}. The color intensity maps to the redshift of the sample: lighter shades represent low redshift, and darker shades represent high redshift.  
For each higher-order bias parameter,
we also plot as a dashed line some relevant best-fit measurements from the literature \citep{2016JCAP...02..018L, 2021JCAP...08..029B, 2022MNRAS.514.5443Z}, which we explain in more detail in the subsequent sections for each parameter.

There are several interesting trends in this result. 
In each panel of Figure \ref{fig:bias}, we see that the central values of the linear bias $b_1$ parameter measurements on the $x-$axis range from $b_1\approx[-0.2,1.9]$. The lower end of that range, from $b_1\approx[-0.2,0.5]$, is primarily occupied by low-redshift, star-forming galaxies (light, blue points), while the higher end of that range, from $b_1\approx[1.5,1.9]$, is dominated by high-redshift, quenched galaxies (dark, red points). This is partly because quenched galaxies are hosted in more massive \citep{2013MNRAS.428.3306W}, and therefore, more strongly clustered halos than their star-forming counterparts. As a result, they have a higher linear bias $b_1$ measurement \citep{2023MNRAS.524.2579B}. 
Meanwhile, the trend in redshift is due to the fact that the linear bias $b_1$ parameter correlates positively not only with halo mass, but also redshift $z$ \citep{1989MNRAS.237.1127C, 1996MNRAS.282..347M};
the same mass halo is a much rarer object at high redshift than today. As all samples are created with roughly similar cuts in stellar mass (see Table \ref{tab:samples} in the Appendix), it is therefore consistent with literature that the higher $b_1$ measurements in Figure \ref{fig:bias} are dominated by high redshift galaxies. We return to these measurements and a comparison to theoretical expectations in \S\ref{sec:theory comparison}, and now focus on the bias relations for the individual parameters shown in Fig.~\ref{fig:bias}.

\subsubsection{\texorpdfstring{$b_2(b_1)$}{b2(b1)}}
\label{subsubsec:b2b1}
Having confirmed that our results in Figure \ref{fig:bias} agree with basic intuition about the behavior of galaxy bias, we examine each bias relation in detail, starting with $b_2(b_1)$ (top left). The central values of the $b_2$ measurements range from $b_2\approx[-0.4, 1.3]$. 
Star-forming galaxies mostly occupy the lower end of that range, from $b_2\approx[-0.4, 0.2]$, although two higher redshift TNG star-forming samples with $b_1 > 1.3$ reach $b_2 \sim 0.5$ and $\sim 1.1$.
In contrast, quenched galaxies occupy a much broader range of parameter space, from $b_2\approx[-0.2,1.3]$. The broadly observed trend is that given a value of $b_1$, quenched galaxies have a higher $b_2$ than star-forming galaxies. In other words, even if quenched and star-forming galaxies have an identical first-order response to changes in large-scale matter overdensities, the former has a larger second-order response. This can be understood as a difference in the shape of the halo occupation distribution for the two galaxy types, which we return to in \ref{sec:theory comparison}.

As we do for the other bias parameters in the following sections, we compare our $b_2(b_1)$ results to several similar measurements in the literature. 
We convert any Eulerian bias parameter measurements into the appropriate Lagrangian equivalent using the co-evolution relations between Eulerian and Lagrangian bias~\citep{2016JCAP...02..018L, Abidi_2018}. We also separate our discussion into analyses of halos and analyses on galaxies. For the former, we plot \cite{2016JCAP...02..018L}'s cubic polynomial fit of the $b_2(b_1)$ bias relation for dark matter halos, which are split by mass (black dashed line). This measurement employs separate universe simulations to measure Eulerian local-in-matter (LIMD) bias parameters.
Although dark matter halos also trace dark matter, their bias is closely related to, but not equivalent to, that of galaxies. As we show later in Equation \ref{eq:PBS}, galaxy bias is an integral of halo bias, the halo mass function, and the halo occupation distribution. Given that we measure the bias relations for galaxies and not dark matter halos in this study, it is consistent with theoretical predictions that our values are different compared to \cite{2016JCAP...02..018L}; this is also what \cite{2022MNRAS.514.5443Z} find in measuring the second-order Lagrangian bias $b_2$ for galaxies. This is analytically understood from Equation~\ref{eq:PBS}, as the HOD upweights higher mass haloes compared to the halo $b_1$ that matches the galaxy $b_1$. The relation between galaxy and halo bias is further explored in \S\ref{sec:theory comparison}.

The other two best-fit relations come from galaxies and are more directly comparable to our measurements. First is a best-fit relation in which \cite{2022MNRAS.514.5443Z} measure the Lagrangian bias relations (pink dashed-dotted line) for galaxies selected using the SubHalo Abundance Matching extended algorithm (SHAMe). The authors set priors using a hypervolume that encloses all of their galaxy samples (pink-shaded region). These priors are noticeably larger than our measurements. This is not surprising; the SHAMe parameters picked in \cite{2022MNRAS.514.5443Z} correspond to galaxy--halo connections that are much broader than what is consistent with known data on galaxy--halo connections\footnote{For example, their Latin hypercube samples power-law slopes for the ${\rm SFR}(M_h)$ relation that are in significant disagreement with what has been constrained from UniverseMachine.}.  
Thus, our tighter constraints are not necessarily a reflection of the model performance but of the realistic types of galaxies we have selected, i.e., quenched and star-forming populations that mimic DESI LRG and ELG targets. 

The second galaxy bias measurement shown in Figure \ref{fig:bias} is a best-fit relation for Eulerian bias parameters determined using field-level forward models of galaxy clustering (grey solid line). With this method, \cite{2021JCAP...08..029B} fit a cubic polynomial to the measured bias relation of TNG galaxies
(grey solid line). To create a Gaussian prior, this fit is assigned a $b_1$-independent $1\sigma$ standard deviation (grey shaded region). Our measurements agree with the general shape of these $b_2(b_1)$ fits but do not exactly match either of them. For instance, in comparison to both of these fits, our measurements of $b_2$ are lower at lower $b_1$, and higher at higher $b_1$.

\subsubsection{\texorpdfstring{$b_{s^2}(b_1)$}{bs2(b1)}}
\label{sec:bs2}

In the top right panel of Figure \ref{fig:bias}, we plot our measurements of the 
$b_{s^2}(b_1)$ bias relation,
which ranges in its central value from $b_{s^2}\approx[-0.3,0.4]$. We show in comparison the Lagrangian local-in-matter (LLIMD) prediction, which assumes $b_{s^2}=0$ (black dashed line). The LLIMD ansatz assumes that the impact of the tidal field on the Lagrangian-space initial positions is negligible
\citep{2016JCAP...02..018L, 2018JCAP...09..008L}. This is an approximation which, while qualitatively accurate, does not strictly hold. Multiple measurements in the literature have found signatures of a nonzero halo tidal bias parameter, and overwhelmingly find that $b^L_{s^2}<0$ for mass-selected halos
\citep{2014PhRvD..90l3522S, 2017MNRAS.472.3959M, 2018JCAP...09..008L}. 
We include in Figure \ref{fig:bias} two measurements from the literature that contradict the LLIMD prediction in galaxies. The first is the best-fit measurement of Lagrangian bias for SHAMe galaxies from \cite{2022MNRAS.514.5443Z} (pink dashed line and shaded region), and the second is the best-fit measurement of Eulerian bias parameters for TNG galaxies from \cite{2021JCAP...08..029B} (grey solid line and shaded region). 
In Figure \ref{fig:bias} we see clear broad trends: the star-forming sample exhibits mildly positive $b_{s^2}$ with no strong scaling with the underlying $b_1$, while the quenched galaxies have negative tidal bias that increases in amplitude with $b_1$. 
It is perhaps surprising that the galaxy types appear to be split by the LLIMD approximation $b_{s^2}=0$, as the positive tidal bias in star-forming galaxies is in contrast to the aforementioned negatively biased measurements of $b_{s^2}$ from halos in the literature. It has been noted previously that spin-dependent assembly bias could push $b_{s^2}$ to positive values in halos~\citep{2021JCAP...10..063L}. In this case, a positive $b_{s^2}$ in star-forming galaxies could be a clear signal of assembly bias in their formation. However, we discuss in \S\ref{sec:assembly}, and show in Figure~\ref{fig:bias_bias} that removing assembly bias in the star-forming galaxy samples has a mild impact on $b_{s^2}$. We comment more broadly on the potential impact of assembly bias on our other parameters in \S\ref{sec:assembly}.

\subsubsection{\texorpdfstring{$b_{\nabla^2}(b_1)$}{bnabla2(b1)}}
\label{sec:bnabla2}

We plot the bias relation for $b_{\nabla^2}(b_1)$ in the bottom left panel. The $\nabla^2\delta$ field is a counter-term in the pure EFT bias expansion of halos. Still, physical intuition can be gleaned from considering that the Laplacian of the Lagrangian density field around peaks corresponds to their curvature. That is, highly concentrated peaks are associated with regions with large $\nabla^2 \delta_L$. Then, the $b_{\nabla^2}$ parameter tells us how haloes form not just in \emph{matter overdensities}, but also in overdensities of varying sharpness given a similarly sized overdensity. In the case of dark matter halos, the associated bias $b_{\nabla^2}$ is expected to be on the order of the squared Lagrangian radius $R_L$ of halos such that $|b_{\nabla^2}(b_1)| \sim R_L^2$\citep{2018JCAP...09..008L, 2019JCAP...11..041L,2022MNRAS.514.5443Z}. Hydrodynamical effects that re-distribute the distribution of matter on one-halo scales are therefore also expected to affect the inferred value of $b_{\nabla^2}$.
It ranges in its central value from $b_{\nabla^2}\approx[-9, 4]$, and is by far the least constrained of the five bias parameters. The $b_{\nabla^2}$ bias parameter consequently has the largest error bars, as the $\mathcal{O}_{\nabla^2}$ field only has appreciable contributions at scales comparable to our filtering scale of $k_\mathrm{max}=0.2 \ihmpc$. The error bars thus drop dramatically as a function of $k_\mathrm{max}$, as Figure \ref{fig:b1b2} shows.

For comparison, we include the best-fit measurement of Lagrangian bias parameters for SHAMe galaxies by \cite{2022MNRAS.514.5443Z} (pink dashed line). Their priors are enclosed in the pink-shaded region. We find that our measurements for $b_{\nabla^2}$ do not show a dependence on the galaxy population type (quenched, star-forming, $M_*$-selected). This contrasts with the other four bias parameters, which give distinct measurements depending on the type of galaxy distribution. 
There are many reasons why this could be the case. The $\mathcal{O}_{\nabla^2}$ operator is sensitive to the curvature of the field around its peaks but is also related to small-scale one-halo effects that affect the inner distribution of matter in a halo. Baryonic feedback, tidal stripping, assembly bias (the effect of which is explored in \S\ref{sec:assembly}), and even halo-finding effects could presumably influence $b_{\nabla^2}$ for a population of halos with similar values of other biases. Indeed, the $b_{\nabla^2}$ parameter is a counter-term that absorbs the sensitivity of the bias expansion to these degenerate small-scale physics.

\subsubsection{\texorpdfstring{$b_3(b_1)$}{b3(b1)}}
\label{sec:b3}

In the bottom right panel, we plot the final bias relation for $b_3(b_1)$, which measures the response of the tracer number density to third-order changes in the dark matter density field.
Our measurements of the cubic bias parameter range from $b_3\approx[-0.4,0.4]$ in central value, with quenched galaxies occupying the higher end of that range (biased) and star-forming galaxies the lower end (anti-biased). The galaxy populations are thus split by $b_3=0$, like the $b_{s^2}$ measurements. This has the effect that
quenched galaxies are more positively biased than star-forming galaxies at a given $b_1$, similar to the $b_2(b_1)$ relation.

We include one comparison to literature in the figure, given by the fitting function from \cite{2016JCAP...02..018L} (black dashed line). This fit remains negative even though our measurements become more positive for higher $b_1$ values. These larger values are especially true for the quenched galaxies, while the star-forming samples
better follow the $b_3(b_1)$ fit: they remain anti-biased, aside from one outlier in TNG at high $b_1$. 

However, it should be noted that we are measuring some ``effective" cubic bias parameter rather than the actual $b_3$. This is because our measurement of $b_3$ includes any error or deviations due to missing higher-order operators, as our model includes only a finite number of bias parameters. Although the effects should be slight, these residual differences could contribute to the consistently larger bias parameter values that we measure. The $b_3$ relation from \cite{2016JCAP...02..018L} also only applies to dark matter halos, and differences between galaxy and halo bias arising from the HOD are exacerbated for a bias parameter that scales more aggressively with halo mass, as is the case for the halo $b_3$ compared to $b_1$ or $b_2$.

Overall, for each bias relation, we see that our measurements 
largely follow the trend lines established by previous best-fit measurements in the literature. We also find that quenched and star-forming galaxies occupy different parts of parameter space for each of the bias relations, aside from $b_{\nabla}^2(b_1)$. The bias for the $M_*$-selected population falls in between the other two measurements at a given number density cut, redshift, and galaxy formation model. Using the terminology presented in \S\ref{sec:assembly}, this is explained by recognizing that the $M_*$-selected HOD is a linear combination of the quenched and star-forming samples.

\subsection{Impact of galaxy formation model}
\label{sec:galaxy model comparison}

\begin{figure*}
    \centering  \includegraphics[width=\textwidth]{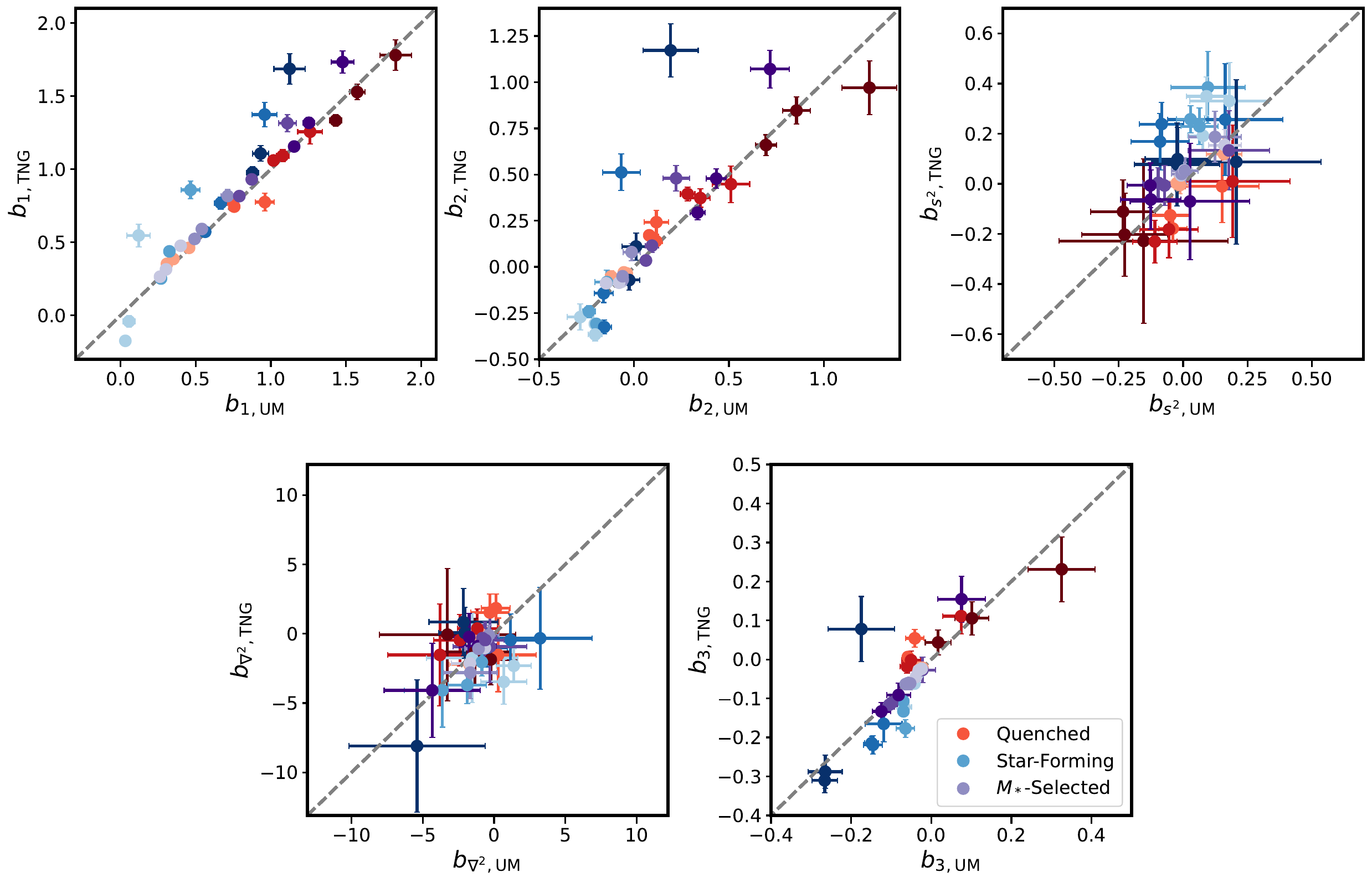}
    \caption{The bias parameter measurements of TNG galaxies in comparison to UM galaxies for quenched (red), star-forming (blue), and $M_*$-selected samples. The color intensity (light to dark) maps to the redshift (low to high).  
    A 1:1 dashed line is plotted for reference.}
    \label{fig:bias_bias}
\end{figure*}

We further investigate how the measurements vary as a function of galaxy formation model by comparing each measurement between UM and TNG directly, shown in Figure \ref{fig:bias_bias}. Here, we plot the bias parameters measured in TNG against those measured in UM at the same redshift ($z=\{0.0, 0.5, 1.0, 1.5\}$), number density (low, medium, high), and galaxy selection type (quenched, star-forming, $M_*$-selected). Looking at the samples collectively, we find that TNG and UM generally measure comparable values for each bias parameter, as there is no significant deviation from the 1:1 line. However, we find larger scatter in $b_{s^2}$ and $b_{\nabla^2}$ as they are less well-measured than the $b_1$, $b_2$, and $b_3$ parameters. 

Interestingly, the populations that deviate from the 1:1 relations in $b_1$, $b_2$, and $b_3$ belong to the lowest number density samples. This suggests that TNG and UM agree reasonably well further down the stellar mass function but disagree on the bias behavior of the most massive galaxies. This could be for several reasons: for example, the exponentially suppressed high-mass tail of the halo mass function may be easily affected by different galaxy formation prescriptions, as it is less constrained by current data and more sensitive to uncertain feedback; the discrepancies for rare samples may also be impacted by the limited box size of TNG. We also note that this increase in scatter with decreasing number density likely signifies that it would be more worrisome to generalize the results and priors presented in \S\ref{sec:priors} to surveys with significantly rarer samples (i.e. lower number density) than those included in our study.

Despite a broader agreement when considering the samples as a whole, we also find that these outliers are
primarily star-forming galaxies, especially in the linear bias $b_1$, a trend which can also be seen in Figures \ref{fig:b1b2} and \ref{fig:Pgm_Pmm} in Appendix~\ref{app:consistency}. This suggests that there is a wider scatter in that population than in the quenched sample; this is likely tied to differences in the modeling approaches between UM and TNG. For example, since UM is an empirical model highly constrained by data, \cite{2019MNRAS.488.3143B} state that the model would benefit from including more constraints on SFRs in massive galaxies at $z \sim 1 - 2$, which is where we see the most deviation from the 1:1 relation in Figure \ref{fig:bias_bias}. In TNG, a hydrodynamical simulation which numerically approximates star formation and feedback, \cite{nelson2021illustristng} note that there are possible observational tensions in the high-mass end of the quenched fraction and galaxy stellar mass relation; these discrepancies may suggest that important adjustments are required in the feedback mechanisms regulating star formation for massive galaxies \citep{2019MNRAS.485.4817D}, and may be linked to the disagreements with UM that we see in our results. 

Finally, we notice a further difference between the models when considering the behavior of the quenched and star-forming samples separately, specifically in the case of the tidal bias $b_{s^2}$. As discussed in \S\ref{sec:bs2}, we find that the quenched galaxies are generally more negatively biased in $b_{s^2}$ than the star-forming galaxies, which tend to be positively biased; however, Figure \ref{fig:bias_bias} shows that these positively biased star-forming galaxies have comparatively higher bias in TNG than their counterparts in UM. The converse is true of the quenched galaxies, which have lower bias in TNG compared to UM, albeit to a lesser extent. 
Combining these two effects means that the aforementioned differences between the tidal bias $b_{s^2}$ of quenched and star-forming galaxies hold for each model but are more pronounced in TNG than in UM. These trends are more readily apparent in Figure \ref{fig:quenched_starforming} of Appendix \ref{app:galaxy type}, which directly compares the bias parameters between the galaxy types for each model. However, while it may be worth further examining this disagreement in a larger box that can better constrain $b_s^2$, the tidal bias likely does not have a large enough impact on the power spectrum for these differences to be detected observationally, unless higher-order information such as the bispectrum is included.

\section{Discussion}
\label{sec:discussion}

In this section, we analyze the results presented in \S\ref{sec:results}. We first discuss the impact of assembly bias in \S\ref{sec:assembly} and follow this up by comparing our bias parameter measurements to theoretically motivated predictions in \S\ref{sec:theory comparison}.
Finally, we examine the degree of non-Poisson stochasticity in our samples in \S\ref{sec:stochasticity}, and end the discussion by setting priors on the bias parameters in \S\ref{sec:priors}. 

\subsection{Assembly bias signature}
\label{sec:assembly}

Halo assembly bias refers to the dependence of halo clustering on secondary halo properties, such as assembly history, in addition to the primary property of halo mass \citep{2005MNRAS.363L..66G, 2006ApJ...652...71W, 2008ApJ...687...12D, 
2020MNRAS.493.4763M}. We are interested in the impact of halo assembly bias on \textit{galaxy} clustering \citep{2006ApJ...639L...5Z, 2007MNRAS.374.1303C}
and in particular, its effects on our measurements of the galaxy bias parameters. We refer to this phenomenon interchangeably as both galaxy assembly bias and assembly bias. 

Unlike simple parameterizations such as HOD models, which explicitly populate halos with galaxies only as a function of halo mass \citep{2001ApJ...546...20S, 2002ApJ...575..587B}, hydrodynamical simulations like TNG and empirical models like UM that are based on full halo formation histories automatically encode galaxy assembly bias into their modeling approach \citep{2018MNRAS.480.3978A, 2019MNRAS.490.5693B, 2021MNRAS.508..698H, 2022MNRAS.512.5793Y}. 
However, the true extent of galaxy assembly bias present in specific 
galaxy samples 
is unknown 
\citep{2013MNRAS.433..515W, 2016MNRAS.460.2552H, 2019MNRAS.485.1196Z}.

For our study to properly inform future clustering analyses on the full range of uncertainty in galaxy formation physics, it is thus crucial that we consider models both with and without assembly bias \citep{2014MNRAS.443.3044Z}. This is especially important when considering separate galaxy population types, such as the quenched and star-forming populations that we examine in this study, 
since such samples are expected to be impacted by assembly bias to different degrees \citep{2007MNRAS.374.1303C, 2021MNRAS.502.3599H, 2022MNRAS.512.5793Y}.
We now present our method of removing assembly bias from each of our samples, as well as its subsequent impact on their bias parameter measurements. 

To remove assembly bias, we follow the standard HOD assumption that the occupation of halos by central $\left\langle N_{\mathrm{cen}}\right\rangle$ and satellite galaxies $\left\langle N_{\mathrm{sat}}\right\rangle$ are two independent processes:
\begin{equation}
    \langle N(M)\rangle=\left\langle N_{\mathrm{cen}}\right\rangle+\left\langle N_{\mathrm{sat}}\right\rangle,
\end{equation}
where $\langle N(M)\rangle$ is the HOD, or mean number of galaxies at a halo mass $M$, $\left\langle N_{\mathrm{cen}}\right\rangle$  is the mean number of central galaxies and $\left\langle N_{\mathrm{sat}}\right\rangle$ represents the mean number of satellite galaxies. We base our subsequent technique on a standard shuffling method in which the galaxy occupation within each halo is randomly exchanged with that of another halo in the same mass bin \citep{2007MNRAS.374.1303C, 2019MNRAS.484.1133C, 2021MNRAS.502.3242X}. We follow a slightly modified procedure, following the example of \cite{2022MNRAS.512.5793Y}: 

\begin{enumerate}
    \item Bin the dark matter halos by their mass from 1$\times10^{10}$ to 1$\times10^{15}$ M$\sun$/h. Measure the central and satellite galaxy occupation $\left\langle N_{\mathrm{cen}}\right\rangle$ and $\left\langle N_{\mathrm{sat}}\right\rangle$ in each mass bin\footnote{We test that the number of halo mass bins does not change this outcome.}.
    \item Given the mass bin of each halo, perform a Bernoulli draw using $\left\langle N_{\mathrm{cen}}\right\rangle$ to determine whether it is populated with a central galaxy. If so, assign that central galaxy the same position as its host halo.
    \item For each satellite galaxy, calculate the offset of its position from the position of its host halo. Construct a catalog of satellite offsets binned by halo mass.
    \item Iterate through each central halo, and perform a Poisson draw using $\left\langle N_{\mathrm{sat}}\right\rangle$ to determine the number of satellite galaxies it hosts. Assign positions to those satellite galaxies by randomly selecting offsets from halos in the same mass bin. 
\end{enumerate}

As with the standard shuffling method, this technique eliminates any dependence of the HOD on any other property besides halo mass. Using the definition of galaxy assembly bias set forth by \cite{doi:10.1146/annurev-astro-081817-051756}, this removes assembly bias from our samples, while by construction preserving the HOD $\langle N(M)\rangle$ of each galaxy sample, as well as $\left\langle N_{\mathrm{cen}}\right\rangle$ and $\left\langle N_{\mathrm{sat}}\right\rangle$. Still, there are two key differences between our procedure and the standard shuffling technique. 

First, we populate the halos by randomly drawing from a probability distribution rather than swapping existing galaxies between halos. Because each dark matter halo has either one or zero central galaxies, we assign central galaxies to halos using the Bernoulli distribution and model the satellite occupation using the Poisson distribution \citep{2005ApJ...633..791Z}. 
However, this means that the number density of galaxies $n_g$ in the shuffled samples is not exactly preserved, as there is a very small scatter around $n_g$ due to the Poisson and Bernoulli variance. 

Second, by populating central and satellite galaxies separately, rather than moving satellites alongside their centrals, we ensure that we also wipe out galactic conformity. This is a phenomenon in which the characteristics of the central galaxy influence the characteristics of its surrounding satellites, e.g., quenched centrals are more likely to be surrounded by quenched satellites \citep{2006MNRAS.366....2W, 2013MNRAS.430.1447K, 2016ApJ...817....9K}. It is important to shuffle the centrals and galaxies separately so that this correlation is eliminated, and the only factor influencing $\left\langle N_{\mathrm{cen}}\right\rangle$ and $\left\langle N_{\mathrm{sat}}\right\rangle$ is the halo mass $M$. 

\begin{figure*}
    \centering
    \includegraphics[width=\textwidth]{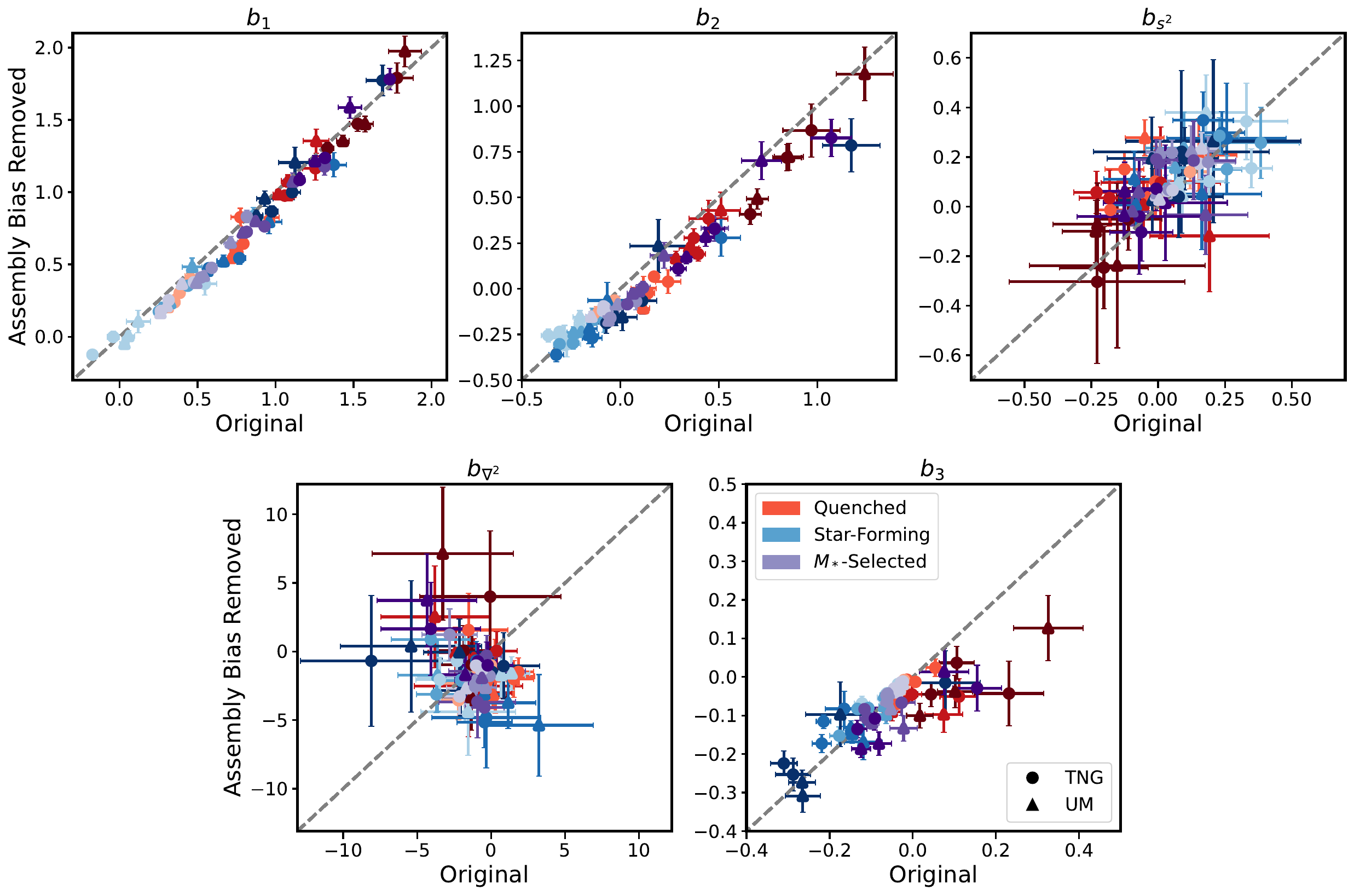}
    \caption{
    The effect of AB removal on each of the bias parameter measurements. The new measurements without AB are plotted against the original measurements with AB at $z=\{0.0, 0.5, 1.0, 1.5\}$ for each sample of galaxies in UM (triangles) and TNG (circles), including quenched (red), star-forming (blue), and the $M_*$-selected population (purple). The color intensity (light to dark) maps to the redshift (low to high). 
    A 1:1 dashed line is plotted for reference.}
    \label{fig:bias_AB_original}
\end{figure*}

We perform this shuffling-based procedure on each of the 72 galaxy samples we created in \S\ref{sec:samples}, and obtain an assembly bias (AB)-removed version of each distribution of galaxies in UM and TNG.  We measure the bias parameters for these AB-removed samples using the methodology of \S\ref{sec:measurement}, and present our results in Figure \ref{fig:bias_AB_original}. Here, we have plotted the bias parameter measurements for each of the AB-removed samples against the measurements from our original samples with assembly bias. The quenched galaxies are plotted in red, the star-forming in blue, and the $M_*$-selected galaxies in purple.

For powers of the matter density field, i.e., the LIMD parameters $b_1$, $b_2$, and $b_3$, removing assembly bias consistently reduces the bias parameter measurements. This is consistent with findings by \cite{2023MNRAS.524.2507H} that both ELGs and LRGs exhibit signatures of assembly bias in which the AB-removed samples are less clustered than the original samples. We find that this effect is stronger for higher-order powers and for higher redshift samples. 

In contrast, there is no discernible trend with the non-local tidal bias parameter $b_{s^2}$, while the $b_{\nabla^2}$ term interestingly shows an inversion of its original values: negative measurements become positive with AB removal, and positive measurements become negative. This effect slightly differs between galaxy populations, such that quenched galaxies become more positive than star-forming galaxies post AB-removal. As a result, quenched and star-forming occupy more distinct parts of AB-removed bias parameter space. This is in contrast to the original $b_{\nabla^2}$
measurements, where there is no apparent pattern, as noted in \S\ref{sec:bnabla2}. 

As defined in \cite{doi:10.1146/annurev-astro-081817-051756}, galaxy assembly bias is caused by halo assembly bias \textit{in addition to} a dependence of the number of galaxies within dark matter halos on a secondary halo property besides halo mass. Since we remove any secondary dependence of the halo occupation distribution through our assembly bias-removal procedure, any changes that we see in the original bias parameter measurements compared to the AB-removed bias parameter measurements would mean that we have indeed detected a signature of assembly bias. We thus see evidence for assembly bias in all bias parameters included in our bias expansion model, except for $b_{s^2}$.

However, interpreting the exact root of these changes is not straightforward, as under the halo model \citep{2000MNRAS.318..203S}, 
galaxy bias can be estimated using what is known as analytic or effective bias \citep{1999MNRAS.305L..21B, 2000MNRAS.311..793B}. 
Through this parameterization, the galaxy bias parameters $b_i$ become a function of the mean galaxy number density $\bar{n}_g$, halo mass function (HMF) $\frac{dn}{dM}$, the HOD $\langle N(M)\rangle$, and the halo bias $b_{i, h}(M)$:

\begin{equation}
\label{eq:PBS}
    b_{i}=\frac{1}{\bar{n}_g} \int d M \frac{d n}{d M} b_{i, h}(M)\langle N(M)\rangle.
\end{equation}
The mean number density of galaxies $\bar{n}_g$ is similarly a function of the HMF $\frac{dn}{dM}$ and the HOD $\langle N(M)\rangle$:

\begin{equation}
\label{eq:PBS_ng}
    \bar{n}_g=\int d M \frac{d n}{d M}\langle N(M)\rangle.
\end{equation}

Understanding how the galaxy bias parameters $b_{i,g}$ change with AB removal would therefore require understanding how exactly the mean galaxy number density $\bar{n}_g$, halo mass function (HMF) $\frac{dn}{dM}$, the HOD $\langle N(M)\rangle$, and the halo bias $b_{i, h}(M)$ are each impacted by an additional dependence on a secondary halo property.
For example, there is a known dependence of the HMF on environment, with the most massive halos forming only in the densest environments \citep{10.1111/j.1365-2966.2009.15402.x}. However,
the relationship of the HOD and halo bias with assembly bias has proven to be quite complicated. For instance, \cite{2020MNRAS.493.5506H} find that galaxy clustering from the basic HOD, which is dependent only on halo mass $M$, underpredicts the TNG300-1 galaxy correlation function by $\sim$15\%. Furthermore, the degree to which AB impacts galaxy clustering seems to depend heavily on which secondary property is considered, with galaxy environment making the biggest difference.

Similarly, there are many studies on halo assembly bias in the linear regime, i.e., the dependence of the linear halo bias $b_{1, h}(M)$ on a secondary halo property. For example, \cite{2018MNRAS.475.4411S} and \cite{2019MNRAS.487.1570S} study the relative bias of halo samples binned by a mass-like property and split by a secondary property, such as spin, concentration, and age. The two studies find complex relationships between halo clustering and the different secondary properties considered, as their impact on halo bias may be asymmetric, e.g., for age and concentration, or it may cross over at a certain halo mass $M$, e.g., spin and concentration. It is also debated which secondary property has the most impact, although 
\cite{2018MNRAS.474.5143M} find that halo spin exhibits the strongest impact on halo bias. 

There are also a few studies on the impact of halo assembly bias on 
halo bias beyond the linear density. \cite{2008MNRAS.387..921A} detect signatures of assembly bias up to fourth order in bias using a proxy for halo concentration. Later studies find strong signatures of assembly bias in $b_1$ and $b_2$ for halo concentration \citep{2017MNRAS.468.2984P}, in addition to halo spin and mass
accretion rate \citep{2017JCAP...03..059L}, as well as in $b_1$, $b_2$ and $b_{s^2}$ for halo age, concentration, and spin \citep{2021JCAP...10..063L}.
There are also signatures of assembly bias in $b_1$ and $b_\phi$, the local primordial non-Gaussianity linear bias parameter, for halo formation time \citep{2010JCAP...07..013R}, halo concentration \citep{2023MNRAS.524.1746L}, as well as halo concentration, spin, and sphericity \citep{2023JCAP...01..023L}.

Unlike these studies, which perform a controlled experiment in measuring the degree of halo assembly bias, we cannot disentangle the effect of the HMF, HOD, and halo bias from each other. Although the effect of halo assembly bias on the bias parameters is fairly well established, the same cannot be said of galaxy assembly bias.
In sum, while all three of these ingredients for predicting galaxy bias are known to be impacted by assembly bias, their complex dependence upon secondary halo property, galaxy type, and galaxy model\footnote{Although it is outside the focus of this study, we even find slight differences in how assembly bias impacts our galaxy samples from TNG and UM.} means that it is beyond the scope of this paper to understand the exact cause of the changes
we see in Figure \ref{fig:bias_AB_original}. We
leave a more comprehensive study of the properties that cause this discrepancy for the future. Nevertheless, having 
stripped away these complications,
we are now prepared to compare our AB-removed measurements of the bias parameters to theoretically motivated predictions in \S\ref{sec:theory comparison}.

\subsection{Halo model prediction of analytic galaxy bias}
\label{sec:theory comparison}

Having removed the complicated dependence of galaxy clustering on assembly bias, we are now able to construct a test of our procedure for measuring the bias parameters. To do so,
we compare our AB-removed measurements of the bias parameters to theoretical predictions of
the galaxy bias parameters for 
$b_1$, $b_2$, and $b_3$. 
As introduced in \S\ref{sec:assembly}, this prediction only assumes dependence upon halo mass $M$. As a result, we expect our AB-removed results to match the theoretical prediction, even if the original measurements do not. We examine the LIMD parameters in particular, as the fitting functions for the non-local bias parameters $b_{s^2}$ and $b_\nabla^2$ are less understood. 

 \begin{figure*}
    \centering
    \includegraphics[width=\textwidth]{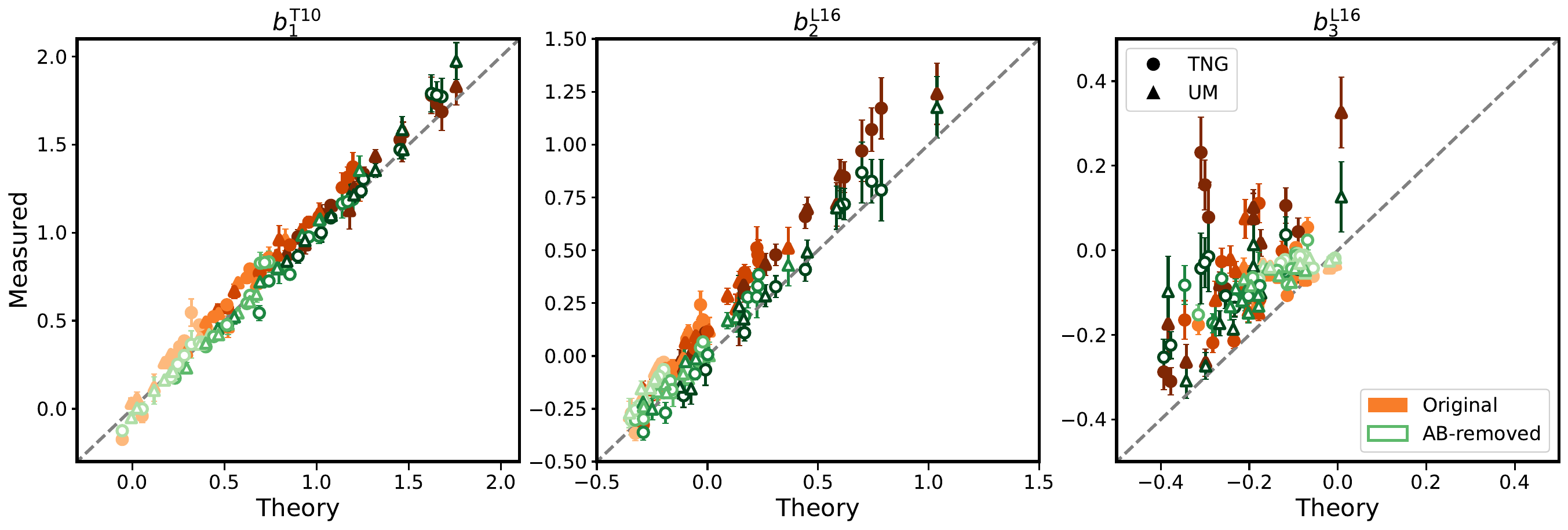}
    \caption{Our measured values of the bias parameters $b_1$, $b_2$, and $b_3$ compared to their theoretically motivated bias predictions for the original (orange) and assembly-bias-removed samples (green).
    The color intensity (light to dark) maps to the redshift (low to high). 
    A 1:1 dashed line is plotted for reference.}
    \label{fig:theory}
\end{figure*}

To make these theoretical predictions, we compute the analytic bias as given in
Equation \ref{eq:PBS}. For $\frac{dn}{dM}$, we use the HMF of \cite{2010ApJ...724..878T}. For $\langle N(M)\rangle$, we empirically derive the HOD by extracting it from each of our TNG and UM galaxy samples. By construction of our shuffling procedure, as detailed in \S\ref{sec:assembly}, this HOD is identical between the AB and AB-removed samples. For the LIMD halo bias parameters $b_{i, h}$, we use theoretically motivated fitting functions.
 First, we calculate the linear halo bias $b_{1,h}$ using the \cite{2010ApJ...724..878T} fitting function:
 \begin{equation}
     b_{1,h}(\nu)=1-A \frac{\nu^a}{\nu^a+\delta_c^a}+B \nu^b+C \nu^c,
 \end{equation}
where $\nu$ is the peak height and the parameters $\{A,a,B,b,C,c\}$ are defined in terms of the overdensity $\Delta=200$. 
 Then, to compute the $b_{2,h}$ and $b_{3,h}$ halo bias parameters, we use the \cite{2016JCAP...02..018L} fitting functions. These are given by
 \begin{equation}
    \begin{split}
    b_{2,h}\left(b_{1,h}\right)= \ &0.412-2.143 b_{1,h}+0.929 b_{1,h}^2+\\
    &0.008 b_{1,h}^3
    \end{split}
 \end{equation}
 and
  \begin{equation}
  \begin{split}
    b_{3,h}\left(b_{1,h}\right)= -&1.028+7.646 b_{1,h}-6.227 b_{1,h}^2+\\
    & 0.912 b_{1,h}^3,
    \end{split}
 \end{equation}
 respectively. We convert all three bias parameters into the Lagrangian frame using the co-evolution relations between Eulerian and Lagrangian bias. Finally, we integrate each component over the mass bins present in our selection. This allows us to make our theoretical prediction of the galaxy bias as a function of halo mass $M$ only, which we compare to our measured bias parameters for the original and AB-removed samples in Figure \ref{fig:theory}. 

Here, we have plotted the original bias parameter measurements (filled orange markers) and the AB-removed bias parameter measurements (open green markers) against our theoretical predictions of the analytic bias. The original bias parameter measurements are larger than the theoretical predictions. This disagreement is stronger for the higher-order parameters but largely vanishes for the AB-removed measurements. This reflects the decreased value of the LIMD parameters on removing assembly bias in Figure \ref{fig:bias_AB_original}. Although the original $b_1$ measurement was in reasonable agreement with the theory, we find overall that the AB-removed bias parameters match the theory predictions better than the original measurements. However, it should be noted that the AB-removed $b_3$ measurement still does not completely match the theory prediction.

To interpret the results of this comparison, we can apply the analytic bias prediction to our measured bias parameters. First we consider the original samples, which do include assembly bias. In this case, it is necessary to use a version of Equation \ref{eq:PBS} that incorporates this dependence. For now, let us simplistically assume that galaxy bias depends on a singular secondary halo property, which we call $X$:

\begin{equation}
\label{eq:PBS_X}
    b_{i}=\frac{1}{\bar{n}_g} \int d M dX\frac{d^2 n}{d MdX} b_{i, h}(M,X)\langle N(M,X)\rangle,
\end{equation}

where

\begin{equation}
\label{eq:PBS_X_n}
    \bar{n}_g=\int d M dX\frac{d^2 n}{d MdX}\langle N(M,X)\rangle.
\end{equation}

In this parameterization, the HMF $\frac{d^2 n}{d MdX}$, HOD $\langle N(M,X)\rangle$, and halo bias $b_{i, h}(M,X)$ each have an additional dependence on some secondary halo property $X$, which does not appear in the original halo mass-only version of Equation \ref{eq:PBS}.
In this case, the original mass function can be written as $\frac{dn}{dM}=\frac{\bar{\rho}}{M}P(M)$, where $\bar{\rho}$ is the average dark matter density and $P(M)$ is the probability of a density fluctuation of mass M collapsing into a halo. By extension, $\frac{d^2n}{dMdX}=\frac{\bar{\rho}}{M}P(M,X)$, where we can use conditional probabilities to write $P(M,X)=P(M|X)P(X)$. 
As a result, simply marginalizing Equation \ref{eq:PBS_X} over the secondary halo property $X$ indeed results in an expression for galaxy bias that is only dependent on halo mass $M$, but 
this marginalized version of $b_i$ is \textit{not} equivalent to the expression given in Equation \ref{eq:PBS}. 
Given this inherent difference, 
it is no surprise that our original bias parameter measurements, which include assembly bias, differ from the theoretical prediction.
This is especially true when considering the more realistic scenario of multiple secondary halo properties affecting galaxy clustering, such as age, spin, and concentration. As discussed in \S\ref{sec:assembly}, it is this complex dependence of the HMF, HOD, and halo bias on various secondary halo properties that, when combined, boost our original bias parameter measurements relative to the theory prediction. 

Now, it is helpful to consider the AB-removed bias parameter measurements:
assuming that they match the theory prediction, this boost can be considered a signature of the strength of assembly bias. This signature appears to be weakest for the linear bias $b_1$ and strongest for the higher order parameters $b_2$ and $b_3$, implying that galaxy assembly bias may be difficult to detect when only considering linear galaxy bias. 
This is contrast to \cite{2021arXiv210112187Z}'s 
findings that $b_1$ is the most affected by assembly bias, even when considering $b_2$ and $b_{s^2}$
in much the same manner as our study: by comparing their results to those obtained from a mass-dependent HOD. 

However, while we find that $b_3$ has a large boost relative to the theoretical prediction, this boost is reduced but not completely removed with removed assembly bias. 
This is particularly noteworthy when considering the $b_3(b_1)$ relation. As discussed in \S\ref{sec:b3}, we find that our measurements of an ``effective" cubic bias parameter are more positively biased than that of \cite{2016JCAP...02..018L}. Although the decrease in $b_3$ upon removing assembly bias cannot fully correct for the upward trend in our original measurements, it is partially mitigated. This could imply that, in addition to the missing cubic bias parameters in our expansion model, assembly bias is part of the reason for the deviation of our original measurements from the dark matter halo $b_3(b_1)$ fit in Figure \ref{fig:bias}.
It also means that it is important to consider what it means for our AB-removed bias parameter measurements to not match the theoretical prediction, as in the case of $b_3$. 

To do this, we can use the form of analytic bias presented in Equation \ref{eq:PBS},
as the only dependence of the AB-removed galaxy bias is on halo mass $M$. First, we consider $\frac{dn}{dM}$: we measure the abundance of halos in TNG and UM and verify that they both 
agree well with the \cite{2010ApJ...724..878T} HMF. 
Second, we consider the HOD $\langle N(M)\rangle$, which has been preserved in the AB-removed samples through our shuffling procedure.
This means that neither the HMF nor the HOD is a source of possible mismatch, leaving the halo bias $b_{i, h}(M)$. 
Any deviation of the AB-removed bias parameter measurements from the theoretical prediction must, therefore, be due to a discrepancy between the halo bias fitting function used and our measured value of the bias parameter. Assuming all fitting functions are correct, Figure \ref{fig:theory} can thus be interpreted as a check of our galaxy bias parameter measurements.

We find that $b_1$ and $b_2$ pass this check, as the AB-removed bias parameters measurements match the theoretical galaxy bias prediction. However, $b_3$ does not pass this check, signifying that our measurement of $b_3$ is not truly the cubic LIMD bias parameter. This is likely due to the fact that we only include one cubic order bias parameter in our bias expansion model, given by Equation \ref{eq:bias expansion}. This choice makes our measurement of $b_3$ a linear combination of the other missing cubic order bias parameters $b_{\delta s^2}$, $b_{s^3}$, and $b_{s t}$, rather than a ``pure" measurement of $b_3$ only.\footnote{It should be noted that while our measurement of $b_3$ may not be accurate, adding it to our model provides an advantage: we ensure that $b_3$ is not included in our definition of the stochasticity field $\epsilon(\bf{x})$, allowing us to minimize $P_{err}(k)$ further and improve our fit of the other bias parameters. We find that $\bar{n}P_{\rm err}(k)$ is relatively flat and
does not suffer from scale dependence until $k\lesssim0.1$ Mpc$^{-1}$h due to the size of the TNG300-1 Dark box.} In addition to this effect, we must consider the possibility that our bias parameter measurements may not be the optimal solution for minimizing the variance of the stochasticity field $\bar{n}P_{\rm err}(k)$, as defined by Equation \ref{eq:loss}. However, given that the other bias parameters have passed numerous checks, including the analytic bias prediction for $b_1$ and $b_2$ in Figure \ref{fig:theory}, a consistency check of $b_1$ in Figure \ref{fig:Pgm_Pmm}, and the convergence of $k_\mathrm{max}$ for all five bias parameters in Figure \ref{fig:b1b2}, this is unlikely. In the latter two figures, we also confirm that the bias parameters do not run with scale, which could signify a breakdown of the bias parameter model, as discussed in Section \ref{sec:measurement}. Due to these checks, we conclude it
is more probable that the deviation of the AB-removed $b_3$ measurements from the analytic bias prediction is simply a consequence of omitting the other cubic order bias parameters, as discussed in \S\ref{sec:b3}.

\subsection{The degree of non-Poisson stochasticity}
\label{sec:stochasticity}
Having reported the measurements of bias parameters in our suite of samples and simulations, as well as the impact of galaxy assembly bias, we examine the stochastic power spectra $P_{\rm err}(k)$ that result from the stochasticity field ${\epsilon}(\bx)$ for each set of best-fit bias parameters $\hat{b}_i$. To calculate this stochasticity field, we generate a realization of the stochasticity field given by
\begin{equation}
    \hat{\epsilon}(\bx) = \hat{\delta}_g (\bx) - 
    \delta_m(\bx)- \sum_i \hat{b}_i \mathcal{O}_i (\bx).
\end{equation}
We do so for both the original and AB-removed samples and assess the degree to which the dimensionless stochasticity spectrum $\bar{n} P_{\rm err} (k)$ deviates from the Poisson prediction of $\bar{n} P_{\rm err} = 1$. 

In previous work, \cite{Kokron_2022} 
found that for LRGs sampled from three different HODs, across the redshift range $0 \leq z \leq 1$, the deviations from stochasticity were small --- on the order of at most 30\% --- except for a single sample in a single redshift bin.
The samples we have constructed in this study allow us to expand on this discussion.
Specifically, we can now understand whether the stochasticity of a galaxy sample changes significantly depending on galaxy type and whether different galaxy formation models conditioned on reproducing similar samples possess the same stochasticity, as we have seen for the case of galaxy bias in previous sections. \par
The small box size of $L = 205 \mpch$ and lack of independent realizations to average over means that the individual spectra for a sample are noisier than those measured in \cite{Kokron_2022}. Thus, we instead fit a perturbative parametric form to each stochastic spectrum~\citep{2018PhR...733....1D}
\begin{equation}
\label{eqn:ptstoch}
    \bar{n} P_{\rm err}(k) = A_0 + A_1 \left ( \frac{k}{k_h} \right )^2,
\end{equation}
where $A_0$ corresponds to the amplitude of the scale-independent contribution, $A_1$ modulates the first scale-dependent correction to the stochasticity, and $k_h$ is the cut-off scale for this expression of the stochasticity and is generically expected to be related to the characteristic scale of the 1-halo regime of the sample in question. If $A_0 > 1$, the sample's stochasticity is super-Poisson, whereas if $A_0 < 1$, it is sub-Poisson. Mechanisms for the generation of sub-and-super Poisson stochasticity have been explored in-depth in past works, and we refer to them for a review~\citep{Hamaus_2010, Baldauf_2013, Kokron_2022, 2024A&A...689A.253B}. We set $k_h = 0.8\ihmpc$ generically to report dimensionless values for $A_1$ in this work.  \par 
\begin{figure*}
    \centering
    \includegraphics[width=\textwidth]{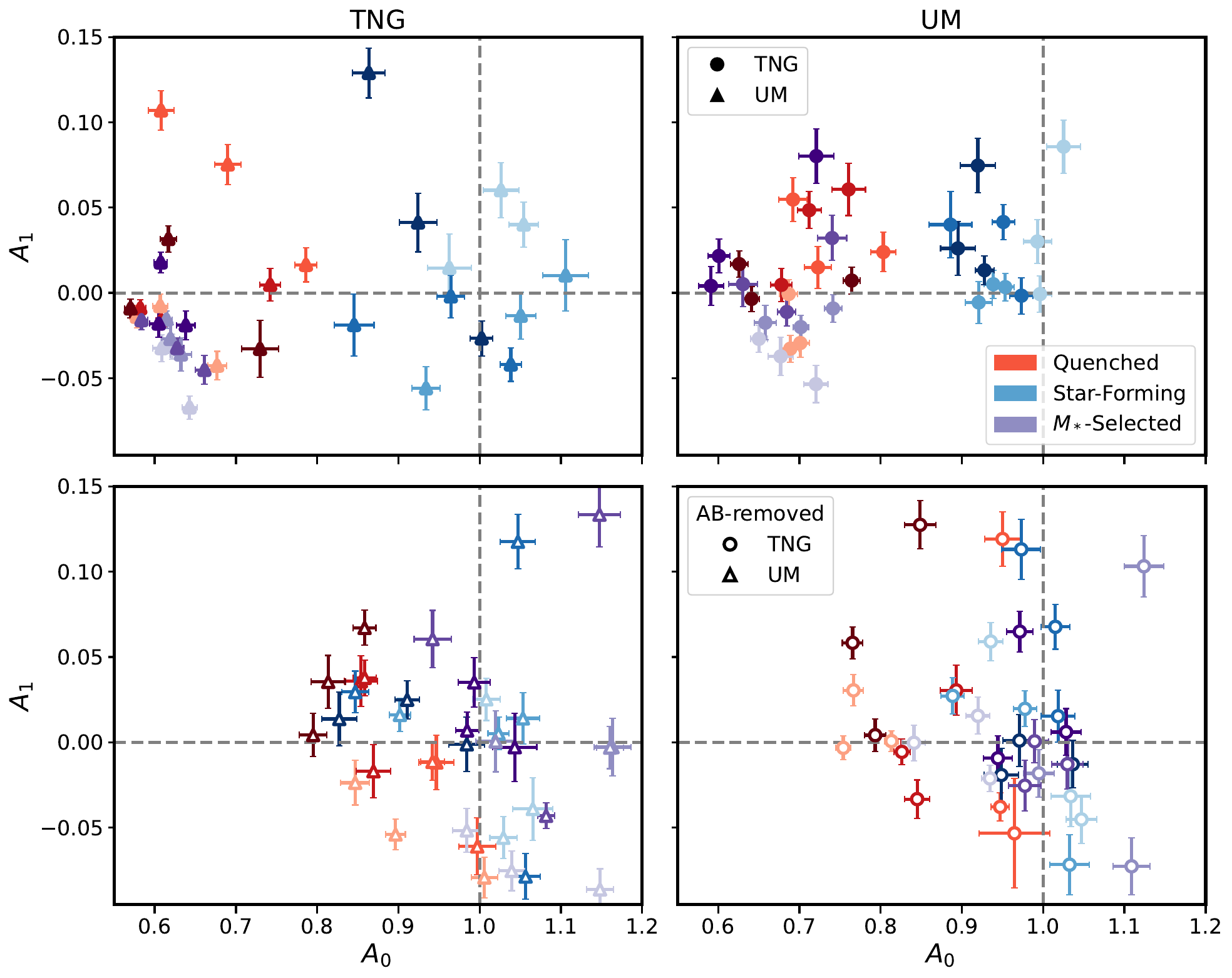}
    \caption{Measurements of the coefficients of the parameterized stochastic power spectrum of the samples considered in this work. $A_0=1$ and $A_1=0$ correspond to a spectrum derived from a Poisson shot-noise distributed sample. The top row shows measurements from TNG and UM, respectively, while the bottom row shows measurements from the versions of these samples with assembly bias removed as described in \S\ref{sec:assembly}. The shading of the color corresponds to varying redshift bins, as in previous figures of this work.}
    \label{fig:stochplot}
\end{figure*}
To measure $A_0$ and $A_1$ for each sample, we fit Equation~\ref{eqn:ptstoch} to the measured stochastic power spectra $\nbar \perr$, assuming equal error-bars per $k$-bin. Using \textsc{SciPy}'s \texttt{curve\_fit} function\footnote{We use the Levenberg-Marquardt method for all applications of \textsc{SciPy}'s \texttt{curve\_fit}.}, we perform the fit within the wavenumbers $0.08 \leq k/\ihmpc < 1.0$, although the results for $A_0$ are broadly insensitive to the exact $k_{\rm max}$ used. Finally, we use the covariance matrix returned by \texttt{curve\_fit} to derive the uncertainty on each measurement of $A_0$ and $A_1$. These results are shown in Figure~\ref{fig:stochplot}, where the measured $A_0$ and $A_1$ are plotted for quenched, star-forming and $M_*$-selected galaxies in both UM and TNG, as well as for the assembly-bias removed samples we define in \S\ref{sec:assembly}.
\par
We first focus on the stochasticity measurements for UM, shown in the upper right panel of Figure~\ref{fig:stochplot}. They reveal a strong bimodality in the distribution of $A_0$ coefficients in the error power spectra of these galaxies, as the star-forming samples have power spectra very close to Poisson-distributed, occupying a narrow range of $\nbar \perr\approx[0.9,1.0]$. This is in contrast to both the $M_*$-selected and quenched samples, which all display strong sub-Poisson $\nbar \perr\approx[0.6 , 0.8]$. The most extreme degree of sub-Poisson stochasticity observed is still consistent with the results of \cite{Kokron_2022}. Although there are some detections of $A_1$ for these samples, the coefficients are very small and tend to be clustered around $A_1 = 0$. Note that due to the $k_h = 0.8 \ihmpc$ normalization we have assumed, the inferred values of $A_1$ correspond to very small corrections to the flat stochasticities we have measured out to one-halo scales. \par 
Contrasting the upper left panel of Figure~\ref{fig:stochplot} with the upper right reveals the differences in stochasticity between the two galaxy formation models we have considered in this work. Namely, we find that for similar galaxy samples, the TNG galaxies possess a larger scatter in both $A_0$ and $A_1$: the star-forming samples, in particular, exhibit a wide range of stochasticity in comparison to UM, while the higher values of $A_1$ in TNG could potentially correspond to the effects of
explicitly implemented baryonic feedback in the stochasticity of the galaxy distribution, for example. 
However, the degree of stochasticity observed is still that of small and controllable deviations from the Poisson expectation, and the bimodality of stochasticity is equivalently apparent in both models. As a result, the measurements for UM and TNG remain consistent with the expectation that field-level stochasticities fit to a suitable bias model will be sub-Poisson for galaxies hosted in more massive halos, due to halo exclusion.
\par 
Finally, the last two panels in the bottom row of Figure~\ref{fig:stochplot} show the impact of removing assembly bias on the stochasticity of galaxy samples. Here, we observe that $A_0$ shifts to values closer to Poisson stochasticity for nearly all of the galaxy samples. This is somewhat intuitive, as the AB-removal procedure explicitly assigns satellite galaxies using a Poisson distribution. However, the collective shift of galaxies towards the Poisson regime points to a potential signature of assembly bias in the small-scale distribution of galaxies within halos. The values of $A_0$ inferred when we remove assembly bias from our samples are also closer in absolute magnitude to those seen in past work, which adopted an HOD to study galaxy stochasticity~\citep{Kokron_2022}. It is interesting that for these quenched samples a prominent impact of removing assembly bias is to reduce the degree of non-Poisson stochasticity. In the standard halo exclusion picture, the effect is solely dependent on the average mass of host halos~\cite{Baldauf_2013}, and so we might expect that the AB-removed HOD would not qualitatively change this stochasticity. \par 
This analysis of quenched galaxies in TNG and UM supports prior conclusions on the distribution of their stochasticity. The extension to star-forming galaxies also reveals that analyses of similar samples, such as ELGs, should be able to use Poisson shot noise as a strong guide for their expected stochasticity. As a last note, we do not find strong evidence for the redshift evolution of $A_0$ or $A_1$ in any of the samples.

\subsection{Priors on \texorpdfstring{$b_i$}{bi}}
\label{sec:priors}

\begin{figure*}
    \centering
    \includegraphics[width=\textwidth]{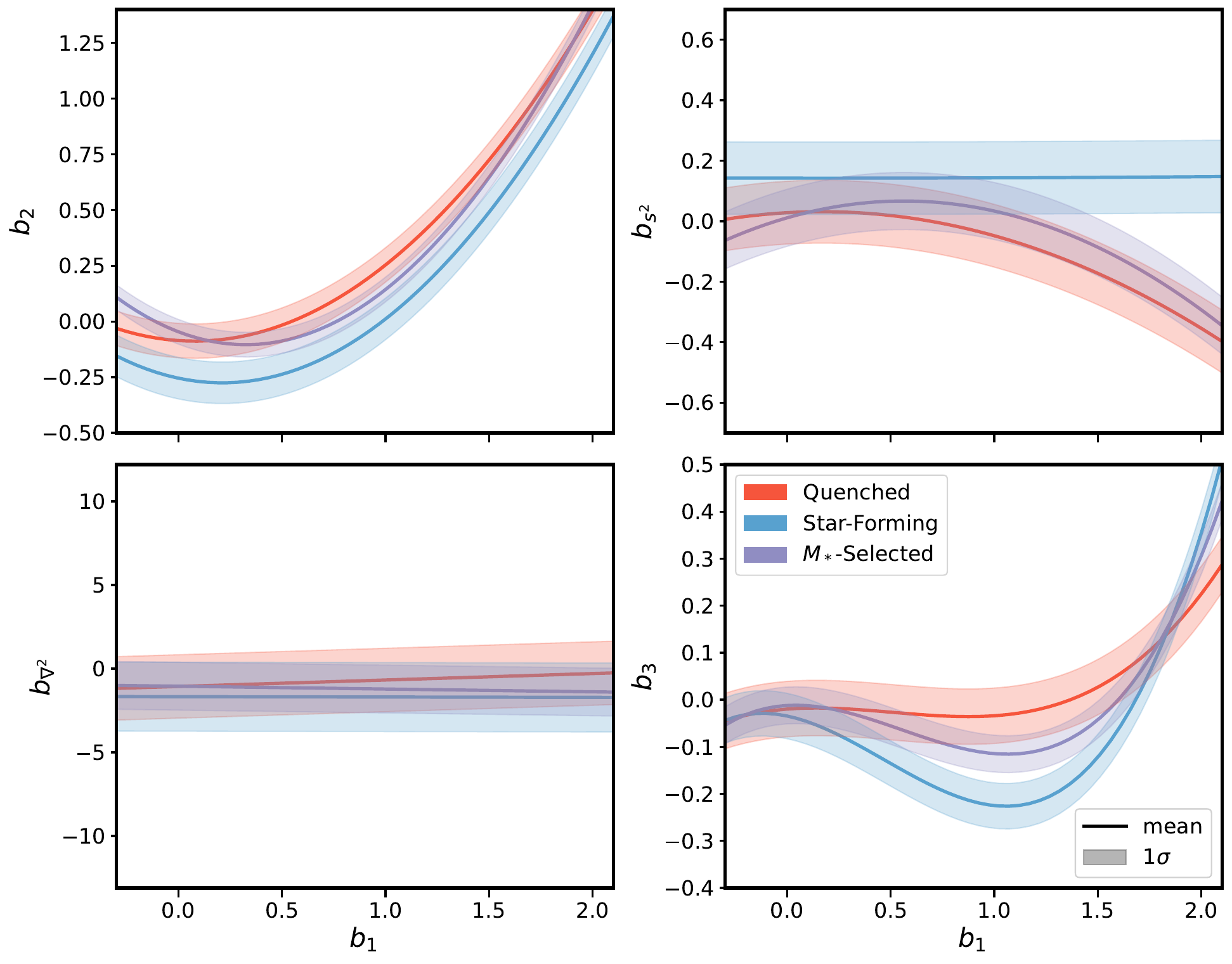}
    \caption{
    Gaussian priors for each of the bias parameter relations as a function of linear bias $b_1$. The bias parameter measurements are plotted for each sample of galaxies with and without AB in UM (triangles) and TNG (circles). Separate priors are set for the quenched (red), star-forming (blue), and the $M_*$-selected population (purple). The 1$\sigma$ standard deviation band is plotted around the best-fit polynomial, as labeled in each panel.}
    \label{fig:priors_bn}
\end{figure*}

We present our priors on the $b_2(b_1)$, $b_{s^2}(b_1)$, $b_{\nabla^2}(b_1)$ and $b_3(b_1)$ bias relations in Figure \ref{fig:priors_bn}. 
We set separate priors for each galaxy population (quenched, star-forming, and $M_*$-selected) and for each bias parameter due to our finding that the measured bias parameters are distinct between different galaxy types. 
In setting our priors, we include samples from both simulations (TNG, UM), all number density cuts (high, medium, low), and each redshift ($z=\{0.0, 0.5, 1.0, 1.5\}$). We include the original and AB-removed galaxy samples to encompass as wide a range of viable galaxy formation physics as possible. This means that 48 galaxy samples are included when setting priors for each galaxy type in Figure \ref{fig:priors_bn}. A detailed plot of the priors alongside these measurements is presented in Figure \ref{fig:priors} of Appendix \ref{app:fullpriors}. 

Although we find outliers in the low-density cuts, as discussed in \S\ref{sec:galaxy model comparison}, 
these samples are, in fact, the most similar to realistic targets of galaxy clustering surveys. It is thus important that we do not exclude any of the galaxy samples and encompass the full range of measured bias parameters in our priors. For this reason, we have chosen to make Gaussian priors around a best-fit polynomial of varying degrees for each galaxy type. The degree of the polynomial fit corresponds to the order of the Lagrangian matter density contrast $\delta_m(\bf{k})$ for each bias field. 
We thus fit a quadratic function to the $b_2(b_1)$ and $b_{s^2}(b_1)$ relations, a line to  the $b_{\nabla^2}(b_1)$ relation, and a third degree polynomial to the $b_3(b_1)$ relation. 

\begin{table*}
    \centering
    \begin{tabular}{llll} 
    \midrule
    \midrule
     $b_i(b_1)$ & Quenched & Star-Forming & $M_*$-Selected \\
     \midrule
    \multirow{2}{*}{$b_2(b_1)$} & $b_2=0.40b_1^2-0.06b_1-0.09$& $b_2=0.46b_1^2-0.20b_1-0.25$& $b_2=0.54b_1^2-0.35b_1-0.05$ \\
     & $\sigma_{b_2}=0.08$ & $\sigma_{b_2}=0.09$ & $\sigma_{b_2}=0.06$ \\
    \midrule
    \multirow{2}{*}{$b_{s^2}(b_1)$} & $b_{s^2}=-0.12b_1^2+0.04b_1+0.03$& $b_{s^2}=0.001b_1^2-0.001b_1+0.14$& $b_{s^2}=-0.17b_1^2+0.20b_1+0.01$ \\
    & $\sigma_{b_{s^2}}=0.10$ & $\sigma_{b_{s^2}}=0.12$ & $\sigma_{b_{s^2}}=0.09$\\
    \midrule
     \multirow{2}{*}{$b_{\nabla^2}(b_1)$} & $b_{\nabla^2}=0.39b_1-1.06$& $b_{\nabla^2}=-0.02b_1-1.67$& $b_{\nabla^2}=-0.17b_1-1.05$\\
     & $\sigma_{b_{\nabla^2}}=1.90$ & $\sigma_{b_{\nabla^2}}=2.06$ & $\sigma_{b_{\nabla^2}}=1.43$\\
    \midrule
    \multirow{2}{*}{$b_3(b_1)$} & $b_3=0.09b_1^3-0.14b_1^2+0.03b_1-0.02$& $b_3=0.24b_1^3-0.34b_1^2-0.09b_1-0.03$& $b_3=0.19b_1^3-0.32b_1^2+0.03b_1-0.01$\\    
    & $\sigma_{b_3}=0.06$ & $\sigma_{b_3}=0.05$ & $\sigma_{b_3}=0.04$\\
    \midrule
    \end{tabular}
    \caption{Gaussian priors on the bias parameter relations $b_2(b_1)$, $b_{s^2}(b_1)$, $b_{\nabla^2}(b_1)$, and $b_3(b_1)$ for quenched, star-forming, and $M_*$-selected galaxies, defined by a best-fit line at $b_i(b_1)$ and the $1\sigma$ standard deviation.
    }
    \label{tab:priors}
\end{table*}

We make this fit using \textsc{SciPy}'s \texttt{curve\_fit} function to carry out a weighted non-linear least-squares analysis. To ensure that our fit is as informed as possible, we take into account that some of our bias parameter measurements are less reliable than others. We do this by incorporating diagonals of the bias parameter covariance matrix, as given by Equation \ref{eq:cov}, on each measured bias parameter. 
We then calculate the standard deviation $\sigma_{b_i}$ of the 48 measured bias parameters from this best-fit line, and 
check separately that the spread of the bias parameters is indeed Gaussian around the best fit. 
The resulting best fit for each bias relation and galaxy type is plotted as a dashed line
in each panel of Figure \ref{fig:priors}, while the standard deviation $\sigma_{b_1}$ is depicted as a shaded region around each best-fit curve.
These best-fit polynomials and Gaussian $1\sigma$ standard deviations are listed for each bias parameter relation and each galaxy type in Table \ref{tab:priors}.

For the priors on the bias parameter relations to be most useful, we also set priors on $b_1$. Given the known evolution of the linear bias $b_1$ with redshift $z$ \citep{2001ApJ...550..522B, 2024JCAP...02..015N}, we choose to set priors by fitting a curve to the time evolution of the linear bias parameter $b_1(z)$ that we measure in our samples. As we do for the bias parameter relations, we use \textsc{SciPy}'s \texttt{curve\_fit} function to carry out a weighted non-linear least-squares analysis. However, to
improve the interpretability of our results, we fit to the following modified function:
\begin{equation}
\begin{split}
    B_1(z)&=\frac{b_1^E(z)}{b_1^E(0)}\\
    &=\frac{1+b_1(z)}{1+b_1(0)}.
\end{split}
\end{equation} 

Rather than simply fitting to $b_1(z)$, we have made two changes here. First, to avoid zero-crossings of $b_1(z)$, we convert our measurement of the Lagrangian linear bias to the Eulerian linear bias via $b_1^E(z)=1+b_1(z).$ Second, to mitigate the vertical scatter due to the varying $b_1^E(z)$ relation among galaxy samples with different number density cuts, we have decided to define the fitting function relative to the value of the Eulerian linear bias $b_1^E(z)$ at $z=0$.

Although the redshift evolution of the first order bias $b_1(z)$ is commonly assumed to be linear in redshift $z$ using a model that conserves the number of galaxies over time \citep{10.1093/mnras/286.1.115}, we choose to model $B_1(z)$ as a quadratic function. Primarily, under peak-background split, the linear bias can be modeled as 
\begin{equation}
    b_1=\frac{\nu^2-1}{\delta_{\mathrm{c}}}
\end{equation}
using the Press--Schechter mass function \citep{1996MNRAS.282..347M}. In an Einstein--de Sitter Universe, the linear growth factor scales as  $D_+(z)\propto(1+z)^{-1}$, which means that we expect the linear bias to scale as $b_1\propto(1+z)^2$ for samples with $\nu(M, z)^2 \gg 1$. This is exactly what we find, as we notice an improved fit to our measurements when modeling $B_1(z)$ as a quadratic, particularly in the $M_*$-selected sample. Additionally, there is a precedent for modeling the redshift evolution of $b_1(z)$ as a second-degree polynomial \citep{2008ApJ...678..627B}. 

To obtain this fit, we thus weight the optimization function for $B_1(z)$ by the inverse of its error, as we do for the prior on the bias relations. However, because $B_1(z)$ depends on both $b_1(z)$ and $b_1(0)$, the error $\sigma_{b_1(z)}$ is not simply equivalent to $\sigma_{b_1(z)}$. Instead, we must propagate the uncertainty of $b_1(z)$ to the new function $B_1 (z)$.
The relative error is thus given by
\begin{equation}
\begin{split}
    \left(\frac{\sigma_{B_1(z)}}{B_1(z)}\right)^2 \approx & \left(\frac{\sigma_{b_1(z)}}{b_1(z)}\right)^2+\left(\frac{\sigma_{b_1(0)}}{b_1(0)}\right)^2+\\
    & 2\left(\frac{\sigma_{b_1(z)}}{b_1(z)}\right)\left(\frac{\sigma_{b_1(0)}}{b_1(0)}\right) \rho_{b_1(z) b_1(0)}\\
    \approx & \left(\frac{\sigma_{b_1(z)}}{b_1(z)}\right)^2+\left(\frac{\sigma_{b_1(0)}}{b_1(0)}\right)^2,
\end{split}
\end{equation}
where we have assumed that the covariance $\rho_{b_1(z) b_1(0)}=0$. 

\begin{figure*}
    \centering
    \includegraphics[width=\textwidth]{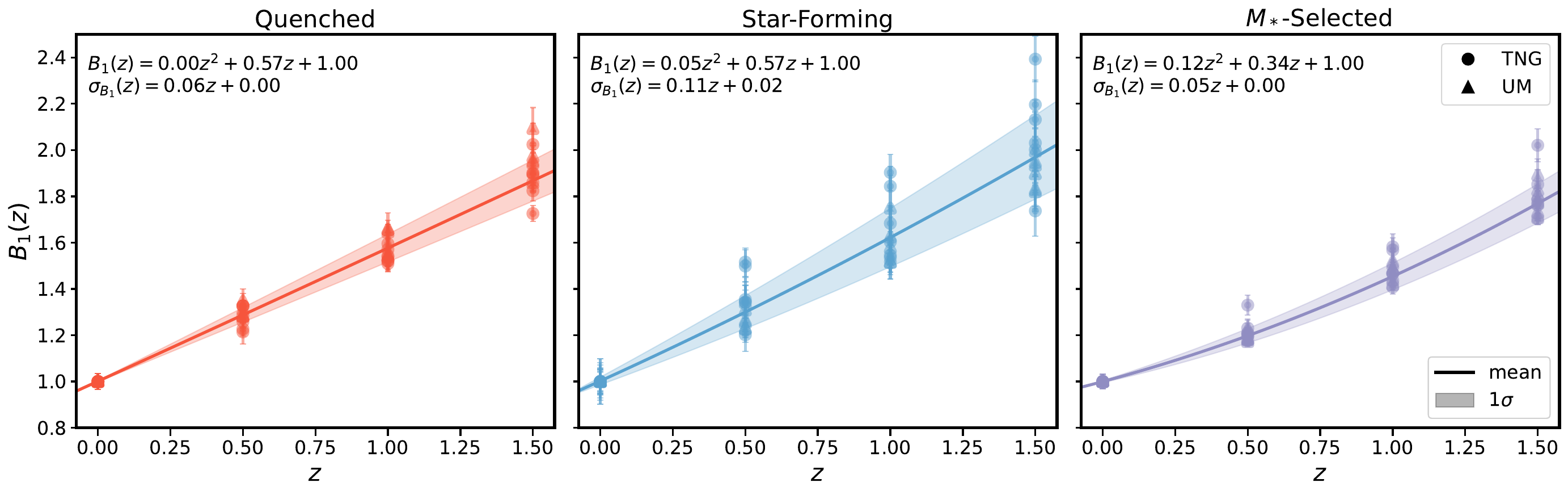}
    \caption{
    Gaussian priors for the time evolution of the linear bias $b_1$ relative to its value at $z=0$, parameterized as a fit to the mean 
    $B_1(z)=\frac{1+b_1(z)}{1+b_1(0)}$. The bias parameter measurements are plotted for each sample of galaxies with and without AB in UM (triangles) and TNG (circles). Separate priors are set for the quenched (red), star-forming (blue), and the $M_*$-selected population (purple). The 1$\sigma$ standard deviation evolves linearly with redshift $z$ and is plotted as a shaded band around the best-fit polynomial, as labeled in each panel.
    }
    \label{fig:priors_b1}
\end{figure*}

Using this method, we present our priors on the linear bias in Figure \ref{fig:priors_b1}, where the best-fit line for $B_1(z)$ is plotted for quenched (red), star-forming (blue), and $M_*$-selected galaxies (purple) separately. Similar to Figure \ref{fig:priors_bn}, each panel thus includes 48 bias parameter measurements across galaxy formation model (UM, TNG), redshift ($z=\{0.0, 0.5, 1.0, 1.5\}$), number density cut (high, medium, low), and assembly bias (original, AB-removed). The modeled $1\sigma$ error of this fit is plotted as a shaded region around each line.

\begin{table*}
    \centering
    \begin{tabular}{lll} 
    \midrule
    \midrule
      Quenched & Star-Forming & $M_*$-Selected \\
     \midrule
    $B_1(z)=0.00z^2+0.57z+1.00$& $B_1(z)=0.05z^2+0.57z+1.00$& $B_1(z)=0.12z^2+0.34z+1.00$ \\
    $\sigma_{B_1}(z)=0.06z+0.00$ & $\sigma_{B_1}(z)=0.11z+0.02$ & $\sigma_{B_1}(z)=0.05z+0.00$ \\
    \midrule
    \end{tabular}
    \caption{Gaussian priors on the linear bias parameter $b_1$, parameterized via the best-fit mean to the function $B_1(z)=\frac{1+b_1(z)}{1+b_1(0)}$ for quenched, star-forming, and $M_*$-selected galaxies. The $1\sigma$ standard deviation evolves linearly as a function of redshift $z$. 
    }
    \label{tab:priors_b1}
\end{table*}

Interestingly, we find that the error $\sigma_{B_1(z)}$ itself evolves as a function of redshift $z$. 
This can be seen directly in Figure \ref{fig:priors_b1}, where the vertical scatter in the bias parameter measurements grows with redshift. 
To parameterize the time evolution of this scatter, we fit a line to the standard deviation of $B_1(z)$ at each redshift $z$, using \textsc{SciPy}'s \texttt{curve\_fit} function to carry out an unweighted linear least-squares analysis. We find that this fit to $\sigma_{B_1(z)}$ describes the measured $\sigma_{B_1(z)}$ quite well, with the largest difference of $\Delta\sigma_{B_1(z)}\approx0.1$ found in the star-forming population; our $\sigma_{B_1(z)}$ fit thus conservatively overestimates the standard deviation of the $B_1(z)$ fit. This fit to $\sigma_{B_1(z)}$, as well as $B_1(z)$, is listed in each panel of Figure \ref{fig:priors_b1}, and is plotted as the standard deviation around the mean. Both fits are summarized in Table \ref{tab:priors_b1}.

We note although that these priors on $b_i(b_1)$ and $B_1(z)$ are based upon cosmology-dependent bias parameter measurements at a singular fixed cosmology, it has been shown previously in the literature that the bias relations are in fact cosmology-independent. Most explicitly, \cite{2024arXiv240910609I} finds that halo mass and cosmology are degenerate, so varying cosmology at fixed halo mass is equivalent to varying halo mass at fixed cosmology, as we have essentially done by varying the number densities of the samples. Intuitively, this also can be understood by the idea that any change in linear bias $b_1$ due to varied cosmology is mirrored by an identical change in the higher order parameters $b_i$, keeping the bias relations $b_i(b_1)$ constant. \cite{2022MNRAS.514.5443Z, 2021JCAP...08..029B} find similar results in their measurements of bias parameters across cosmologies. This means that our priors are robust even in a broader cosmological parameter space, and may be directly adopted in future HEFT-based analyses that use galaxy clustering data to improve constraints on cosmological parameters \citep{2021JCAP...09..020H, 2024arXiv240704795C,2024arXiv240704607S,2024JCAP...02..015N}. 

As we find that the priors on quenched, star-forming, and $M_*$-selected galaxies behave distinctly from one another, we summarize these differences in the following sections. When applicable, we also provide guidance on their interpretation. For those who may prefer more conservative bounds, we include in Appendix \ref{app:fullpriors} a set of priors based on all 144 samples together, rather than separating by galaxy type. These alternative priors on the bias relations $b_i(b_1)$ are presented in Figure \ref{fig:priors_all} and Table \ref{tab:priors_all}, while those on $B_1(z)$ are shown in Figure \ref{fig:priors_b1_all} and Table \ref{tab:priors_b1_all}. These conservative priors may be more appropriate for those seeking to marginalize over an even broader range of galaxy formation models than those considered in our analysis. In such cases, one might also consider increasing the standard deviation of our priors --- while ensuring they remain informative --- as the overall consistency of our bias parameter measurements across models suggests that we can, and perhaps should, take advantage of this robustness.

\subsubsection{Quenched galaxies}
The quenched galaxies are more positively biased in $b_1$ and $b_3$ than the star-forming galaxies at a given $b_1$. They are also slightly more biased in $b_{\nabla^2}(b_1)$ than the star-forming galaxies. However, due to the large error bars, there is so much overlap between the two populations that any differences are negligible. In fact, all galaxy types, including the $M_*$-selected sample, are roughly constant at $b_{\nabla^2}(b_1)\approx-1.$ Finally, although we allow for a second-degree dependence of $B_1(z)$ on redshift $z$, the time evolution of the linear bias for quenched galaxies remains very strongly linear. 

Now, we consider the standard deviation of our priors: we find that the quenched galaxy population has much tighter bounds on all five bias parameters than the star-forming galaxy population. These tighter priors are slightly larger than, but still fairly comparable to the error in the $M_*$-selected sample. This suggests that there is less variation in modeling quenched and $M_*$-selected galaxies than star-forming galaxies for the galaxy formation models we consider.

\subsubsection{Star-forming galaxies}
\label{sec:priors_starforming}

We find that the star-forming galaxies are nearly constant in tidal bias at $b_{s^2}(b_1)\approx0.14$, while the quenched and $M_*$-selected samples follow a complex quadratic function that peaks at $b_1\approx0.5$ and becomes increasingly more anti-biased. For the linear bias $b_1(z)$, we find that the star-forming sample has an even stronger dependence on redshift than the quenched galaxies, as both the slope of $B_1(z)$ and $\sigma_{B_1}(z)$ are larger than that of the star-forming galaxies. It should also be noted that while the error $\sigma_{B_1}(z)$ of star-forming galaxies appears linear, it is slightly quadratic. 

This larger slope in  $\sigma_{B_1}(z)$ means that the error in $B_1(z)$ for star-forming galaxies grows more rapidly with redshift than it does for both the quenched and $M_*$-selected populations. Overall, these wider priors for both the bias relations and time evolution of the linear bias of star-forming galaxies suggest that this galaxy population is the least agreed on by the models we consider and suffers from the most uncertainty in galaxy formation physics.

\subsubsection{\texorpdfstring{$M_*$}{M*}-selected galaxies}
\label{sec:M*-selected}

The $M_*$-selected sample is characterized by properties that smoothly interpolate between the quenched and star-forming samples. This behavior is consistent with the fact that the halo occupation distribution for the $M_*$-selected sample is a linear combination of the quenched and star-forming galaxies. This means that the $M_*$-selected galaxies occupy a part of bias parameter space between the quenched and star-forming samples, a trend most evident in the prior for $b_3(b_1)$. Here, we find that the shape of the prior is very similar to that of the star-forming sample, meaning that the $b_3$ of the $M_*$-selected sample evolves with linear bias $b_1$ in a similar manner to the star-forming sample. However, the former is shifted towards more positive values of $b_3$ than the latter, similar to the quenched sample. 

Interestingly, we find that the $M_*$-selected sample has the tightest priors out of all the galaxy types that we consider. This is most evident in the $b_2(b_1)$ and $b_3(b_1)$ relation, as well as the linear bias. 
It should be noted that the $M_*$-selected sample has twice the number density of the other galaxy populations, due to the threshold-based selection method described in \S\ref{sec:samples}. This may contribute to the tighter priors we measure, given that the covariance and galaxy number density are inversely related, as shown by Equation \ref{eq:cov}. However, due to the additional dependence upon the $M_{ij}$ matrix, this is not a definitive explanation.

These tighter priors for the $M_*$-selected sample suggest that the galaxy formation models we consider agree relatively more on how to model the stellar mass of galaxies than they do their star-formation rate, which we use to distinguish between the quenched and star-forming population. We see this disagreement partly reflected by the threshold cuts in $M_*$ and sSFR used to make the galaxy samples, listed in Table \ref{tab:samples} of the Appendix. The difference in the threshold for sSFR between the two models makes it all the more compelling that the distinct parts of bias parameter space occupied by the different galaxy types are reproduced in both models, showing that the bias expansion is agnostic of the particular galaxy formation model.

Finally, it should be noted that although the quenched and star-forming galaxy populations are well-described by a Gaussian prior, the $M_*$-selected population has some outliers from the lowest number density cut in TNG; this is the same population of outliers noted previously in \S\ref{sec:galaxy model comparison}. Due to the outliers, the $B_1(z)$ fit 
is shifted slightly downwards of the overall $M_*$-selected measurements, meaning that the fit more accurately describes lower values of $B_1(z)$ for this sample.  
This results from the weighted fit that we perform on the bias parameter measurements, which preferentially prioritizes samples with smaller error bars. We see this clearly in Figure \ref{fig:priors_b1}, as the error on $B_1(z)$ is highest for the $M_*$-selected samples that deviate the most from the fit. 

\subsection{Comparison with other work}
\label{sec:note}
As this paper was nearing submission, the work of \cite{ivanov2024millenniumastridgalaxieseffective} was posted to the arXiv. They perform field-level fits in the ``shifted operator basis" and Eulerian bias parameterization of \cite{Schmittfull_2019} to samples of quenched and star-forming galaxies in two hydrodynamical simulations, as well as to explicitly parametric halo occupation distributions for LRG- and ELG-like samples. Where there is overlap, we broadly find that their results are consistent with those presented here. For example, this work also shows a split on  $b_2(b_1)$ for star-forming and quenched galaxies, has consistent results for real-space stochasticities, and shows a split between quenched and star-forming galaxies along the LLIMD prediction for the tidal bias parameter. Our work covers a wider redshift range and more distinct approaches to galaxy formation since, in addition to hydrodynamical simulations, we consider an empirical model tuned to a wide range of data and explicitly test for the impact of assembly bias. This allows us to demonstrate that the bias relations for galaxy populations at fixed number densities are only weakly sensitive to both redshift and to quite different galaxy formation prescriptions. This consistency with our work, despite different simulations and different methodologies for both generating the field-level forward model and fitting associated parameters, corroborates the robustness of our measurements and reported priors. Together, these papers give a coherent picture of the impact of uncertain galaxy formation physics on field-level bias models and provide further support for the importance of careful, analysis-specific work to quantify these uncertainties in the future.

\section{Conclusions}
\label{sec:conclusions}

In this work, we present direct, field-level comparisons of different models of galaxy formation using a second-order hybrid N-body Lagrangian perturbative bias expansion. We fit the bias expansion model to the density fields of galaxies from different models with the same initial conditions, while varying the number density, redshift, and type of galaxy included in our study. By using the same set of initial conditions, we can make a one-to-one comparison between models that include disparate approaches to galaxy formation. This allows us to estimate the range of bias parameters expected for typical galaxy samples over a range of reasonable, physically motivated models for galaxy formation, providing 
an essential ingredient for extracting the most precise cosmological information from current and future 
large galaxy surveys. 

Specifically, we examine the \textsc{UniverseMachine} (UM), an empirical model of galaxy formation, and the \textsc{IllustrisTNG} (TNG) simulation. For each galaxy formation model, we make cuts on stellar mass ($M_*$) and specific star-formation rate to create catalogs of quenched and star-forming galaxies at a range of target number densities and redshifts ($z=\{0.0, 0.5, 1.0, 1.5\}$). We also create a duplicate set of galaxy samples with assembly bias (AB) removed. These AB-removed samples can be considered an extension to the range of possible galaxy formation models. Consequently, by exploring the bias parameter space for a given galaxy population, we can investigate in detail not only the difference in bias between quenched and star-forming galaxies, but also quantify the uncertainty in these parameters due to changes in the galaxy formation model and uncertainties in the galaxy--halo connection. 

Our primary findings are as follows:
\begin{enumerate}
    \item {\bf We characterize the bias relations, which define the higher order bias parameters as a function of the linear bias ($b_2(b_1)$, $b_{s^2}(b_1)$, $b_{\nabla^2}(b_1)$, $b_3(b_1)$), for 72 different galaxy samples (Figure \ref{fig:bias}).} We find general agreement between our bias parameter measurements and those in the literature that look at similar samples. 

    \item {\bf Star-forming and quenched galaxies occupy different parts of parameter space in the bias relations (Figures \ref{fig:bias}, \ref{fig:quenched_starforming}).} 
    At fixed $b_1$, the $b_2$ and $b_3$ of quenched galaxies are larger than those of star-forming galaxies. In contrast, the opposite is generally true of the tidal bias: $b_{s^2} \leq 0$ for quenched galaxies, while $b_{s^2} \geq 0$ for their star-forming counterparts. $b_{\nabla^2}(b_1)$ is the only bias relation for which we find no discernible differences between quenched and star-forming galaxies. 

    \item {\bf There is broad agreement between the bias relations for different models of galaxy formation (Figure \ref{fig:bias_bias}).} Low number density samples, which are currently less constrained by data and subject to more uncertain galaxy formation physics, exhibit more scatter (Figures \ref{fig:bias_bias}, \ref{fig:quenched_starforming}).  There is also a wider range in the measured bias of star-forming galaxies, which generally tend to be more different between the UM and TNG models (Figures \ref{fig:Pgm_Pmm},  \ref{fig:b1b2}).

    \item {\bf We examine 72 additional galaxy samples with realistic levels of assembly bias, and find that they exceed analytic bias predictions that depend only on halo mass (Figure \ref{fig:theory}).}  
 They match these predictions for $b_1$ and $b_2$ after removing assembly bias, upon which $b_2$ and $b_3$  decrease, $b_1$ and $b_{s^2}$ remain unchanged, and $b_{\nabla^2}$ undergoes an orthogonal shift (Figure \ref{fig:bias_AB_original}). This indicates that measured deviations from the analytic bias predictions may be viewed as a potential signature of assembly bias.
 
    \item {\bf We quantify the degree of non-Poisson stochasticity in all 144 galaxy samples (Figure \ref{fig:stochplot}).} We find that star-forming samples are roughly Poisson, while the quenched and $M_*$-selected samples are sub-Poisson ($\bar{n}P_{\mathrm{err}}(k)\approx[0.6,0.8]$); this is expected because the latter are hosted in massive halos and experience more halo exclusion. 
    Interestingly, we find that removing assembly bias shifts these samples from sub-Poisson to Poisson noise. This points to a potential signature of assembly bias in the one-halo regime.
    
    \item {\bf We use our full set of samples to estimate the range of uncertainty in representative models of galaxy formation and set Gaussian priors on the bias relations (Figures \ref{fig:priors_bn}, \ref{fig:priors_b1}, \ref{fig:priors}, \ref{fig:priors_all}, \ref{fig:priors_b1_all}).} For each redshift, number density, and galaxy type (quenched, star-forming, $M_*$-selected), these priors encompass both galaxy formation models (TNG, UM) with and without assembly bias. While this is not a comprehensive marginalization over all possible galaxy formation models, it provides a reasonable assessment of the expected range given current understanding. 
    These priors are not survey-specific, and can be directly used in cosmological analyses that employ Lagrangian Perturbation Theory and Hybrid Effective Field Theory (HEFT).
\end{enumerate}    

The techniques we develop in this work allow us to directly compare any galaxy formation models using the bias expansion, as well as learn more about different galaxy populations. With this methodology
in hand, there are several scientific directions we can now explore further. A natural first step is applying our analysis to other relevant targets of future surveys that require non-linear bias models to reliably extract cosmological parameters and demystify galaxy formation physics. 

Compelling tracer populations include Lyman-break galaxies (LBGs), which are massive, star-forming galaxies selected by their photometric colors that can serve as a probe for the high-redshift ($z\approx[2,5]$) Universe, as well as inform science on dark energy, redshift-space distortions, and neutrino mass constraints \citep{2001ApJ...554...85W, 2016MNRAS.461..176P, 2017MNRAS.469.4428J, 2019JCAP...10..015W, 2024ApJ...961...27N}. A similarly rare, high redshift tracer that would benefit from a robust model of non-linear bias is Lyman-alpha emitters (LAEs): young, star-forming galaxies residing in less massive halos than LBGs at $z\approx[2,6]$ \citep{2015MNRAS.453.1843S, 2024JCAP...06..052E, 2024arXiv240302414R}. These tracers provide exciting avenues forward, especially given the future DESI-II upgrades, which plan to target LBGs and LAEs \citep{2022arXiv220903585S}.

Moving forward, it would be worthwhile to conduct such an analysis in a larger simulation volume, as the TNG300-1 Dark from the \textsc{IllustrisTNG} Project has a rather small box size of $L=205 \ihmpc$. This makes our results noisier than they would be in a larger box, as the error bars on our measurements scale inversely with volume (see Equation \ref{eq:cov}). An example of a more suitable simulation suite is the newest MilleniumTNG Project \citep{2023MNRAS.524.2556H}, which pushes the \textsc{IllustrisTNG} galaxy formation model to a volume that matches the original Millennium simulation \citep{2005Natur.435..629S}. Their largest run, MTNG740, spans a box size of $L=500\ihmpc$: this is nearly a factor of 15 larger in volume than the \textsc{IllustrisTNG} box we have used in this work, meaning that the error bars on our bias parameter measurements would be reduced by a factor of about four. Yet another possibility is the \textsc{FLAMINGO} suite of hydrodynamical simulations \citep{2023MNRAS.526.4978S}, which has dark-matter only counterparts for their boxes of size $L=1$ Gpc and $L=2.8$ Gpc; these would reduce our error bars by a factor of about 6 and 30, respectively.
Such improvements would be particularly valuable for better understanding the noisiest bias parameters, $b_{s^2}$ and $b_\nabla^2$.

An additional avenue for reducing noise and increasing the accuracy of our bias parameter measurements is to investigate the impact of minimizing higher-order statistics of the stochasticity field, as in \cite{2019PhRvD.100d3514S}. Currently, we minimize the field-level variance, as Equation \ref{eq:loss} shows, but it is possible to minimize, for example, the skewness or kurtosis simultaneously. It would be interesting to see the level of precision gained from this additional step.

Finally, although we have only lightly touched on assembly bias in this work, it is clear that it will be important for future galaxy clustering analyses to incorporate a dependence on secondary halo properties in their models of the galaxy--halo connection. This will be crucial for conducting accurate cosmological parameter inference. As a result, it would be interesting for a subsequent study to further investigate the effect of removing assembly bias on galaxy bias parameters, particularly how these changes compare among different models of galaxy formation and different galaxy types. It is especially important to understand how and why these changes occur, i.e., how exactly the halo mass function, halo occupation distribution function, and halo bias are influenced by certain secondary properties; this full study is beyond the scope of this work. A separate but related topic is the degree of non-stochasticity, as discussed in \S\ref{sec:stochasticity}, which invites a full investigation into the impact of assembly bias on the stochasticity of our galaxy samples.

In sum, our findings show that quantifying the range of bias relations for physically motivated models of galaxy formation can provide tightened priors on bias relations, presenting an exciting avenue forward for HEFT-based cosmological constraint analyses, such as the recent work of \cite{2021JCAP...09..020H, 2024arXiv240704795C,2024arXiv240704607S,2024JCAP...02..015N}, as well as inference at the field-level \citep{2024PhRvL.133v1006N}. The estimates provided here are particularly relevant for future DESI galaxy clustering measurements for LRGs and ELGs, as our galaxy selection and number density cuts are intended to mimic those targets. The techniques developed in this work will also be useful for maximizing information from other galaxy clustering data in the coming years, such as LSST, SPHEREx, Euclid, Roman, and beyond. Care should be taken to develop robust priors for such surveys tuned to the specific samples used for cosmological analyses; each sample will have different galaxy formation systematics, and each analysis will have different systematic requirements. 
Although we are not yet able to provide a true marginalization over all plausible models of galaxy formation, the modest range of bias relations that we find from a wide range of physically motivated galaxy formation models emphasizes the following: a robust understanding of galaxy formation can provide essential information that will significantly strengthen cosmological constraints from future surveys.

\section*{Acknowledgments}
We thank Chun-Hao To for making the field-level Lagrangian component fields at $z=0$ available and for helpful discussions, and Stephen Chen, Elisabeth Krause, Philip Mansfield, Kate Storey-Fisher, Sihan Yuan and the GFC group at Stanford for useful discussions and comments. We are grateful to The \textsc{IllustrisTNG} Project for making the simulation suite used here publicly available, and to the Universe Machine team for their public data and code.

This work received support from Stanford University and from the U.S. Department of Energy under contract number DE-AC02-76SF00515 to SLAC National Accelerator Laboratory. MS acknowledges support from the Stanford Graduate Fellowship, the EDGE: Enhancing Diversity in Graduate Education Doctoral Fellowship Program, and from the National Science Foundation Graduate Research Fellowship under Grant No. DGE-2146755. Any opinions, findings, and conclusions or recommendations expressed in this material are those of the authors and do not necessarily reflect the views of the National Science Foundation. NK acknowledges support from NSF award AST-2108126 and the Fund for Natural Sciences of the Institute for Advanced Study. 

This research has made use of NASA’s Astrophysics Data System and the arXiv preprint server, as well as computational resources at SLAC National Accelerator Laboratory and the Sherlock cluster at the Stanford Research Computing Center (SRCC). The authors are thankful for the support of the SLAC computational team and the SRCC team. Some calculations and figures in this work have been made using
nbodykit \citep{2018AJ....156..160H}.

\section*{Data Availability}

All data from the TNG300 simulation are available for download on the \href{https://www.tng-project.org/data/}{\textsc{IllustrisTNG} Project's} website. Access to other data underlying this paper is available upon reasonable request to the authors.

\appendix

\section{Sample selection}
\label{app:sampletable}

\begin{table*}
    \begin{tabular}{llllllll} 
    \midrule
    \midrule
    \multirow{3}{*}{$z$} & \multicolumn{3}{c}{\multirow{2}{*}{Number Density/(10$^{-4}$ $ [\mpch]^{-3}$)}} & \multicolumn{4}{c}{Threshold} \\
       & & & & \multicolumn{2}{l}{Log($M_*$/$M_\odot$)} & \multicolumn{2}{l}{Log(sSFR/(year$^{-1})$)}  \\ 
      \cmidrule{2-8}
     & Quenched & Star-Forming & $M_*$-Selected                                   & TNG & UM                                          & TNG & UM   \\ 
    \midrule
    \T~& 8.3   & 1.7       & 10                                    & 10.8        & 10.9                                                   & -11.7      & -10.7                                                                             \\
     0 & 33 & 6.5 & 39 & 10.4 & 10.5 & -11.3 & -10.3 \\
     \B~& 57   & 11       & 68                                  & 10.2       & 10.3                                                  & -10.2      & -10.1                                                                            \\
    \midrule
    \T~& 5.0     & 5.0        & 10                                    & 10.8       & 10.8                                                  & -11.7 & -11.5                                                                             \\
     0.5 & 20 & 20 & 39 & 10.4 & 10.5 & -11.5 & -11.3 \\
     \B~& 34   & 34      & 68 & 10.1 & 10.3 & -10.4 & -11.2 \\   
    \midrule
    \T~& 5.0     & 5.0        & 10                                    & 10.7       & 10.8                                                   & -10.9 & -11.3                                                                             \\
    1 & 20 & 20 & 39 & 10.3 & 10.4 & -10.2 & -11.0 \\
    \B~& 34   & 34      & 68 & 10.0 & 10.2 & -9.60 & -9.80 \\   
    \midrule
    \T~& 5.0     & 5.0        & 10                                    & 10.6       & 10.7                                                   & -10.3 & -11.0                                                                             \\
     1.5 & 20 & 20 & 39 & 10.1 & 10.2 & -9.40 & -9.40 \\
     \B~& 34   & 34      & 68 & 9.90 & 9.90 & -9.30 & -9.10 \\   
    \midrule
    \end{tabular}
    \caption{The number density of each galaxy population (quenched, star-forming, $M_*$-selected) at each redshift ($z=\{0.0, 0.5, 1.0, 1.5\}$) used in this study, listed in increasing order. The stellar mass and specific star formation rate used to split each sample is also included under the column labeled ``Threshold".}
    \label{tab:samples}
\end{table*}

As discussed in \S\ref{sec:samples}, we select the galaxy samples in this study by making cuts in $M_*$ and sSFR to achieve a certain target number density (low, medium, high) that is identical for each model (UM, TNG). The results of this sample selection are summarized in Table \ref{tab:samples}, where the total number density of each sample is listed under the $M_*$-selected column. This galaxy sample is split evenly between quenched and star-forming galaxies at each redshift $z$, aside from $z=0$, which is split such that the number density of quenched galaxies is five times as large as that of star-forming galaxies\footnote{Refer to \S\ref{sec:samples} for an explanation of this choice.}. 
This split in the $M_*$-selected population is made by selecting a threshold in sSFR, below which a galaxy is considered quenched and above which a galaxy is considered star-forming. The ``Threshold" column of Table \ref{tab:samples} lists the minimum $M_*$ that we include in our samples, as well as the sSFR at which we distinguish between quenched and star-forming galaxies. 

It is noticeable that the $M_*$ cut is very similar between the TNG and UM models, which agree to 0.1 dex. We see more variation in the sSFR thresholds, of up to 1 dex. This variability in the sSFR threshold may be a reflection of the limitations of our straightforward sample selection method, which does not attempt to model galaxy colors or make cuts in color--magnitude space. However, our finding that the bias parameters are distinct between these two populations --- something that is reproduced in both models --- suggests that this is more likely due to a fundamental disagreement in what UM and TNG consider to be quenched and star-forming galaxies. Moreover, we find that the sSFR cuts we make using our $M_*$-based selection procedure does generally match the log(sSFR/[year$^{-1}])=-11$ cut commonly adopted by the literature; this is true for the majority of our samples, which span log(sSFR/[year$^{-1}])\approx[-9.10, -11.7]$, and skew higher only for the highest number density, highest redshift samples.

\section{Consistency with large-scale linear bias and running with \texorpdfstring{$k_{\rm max}$}{kmax}}
\label{app:consistency}
It is important to assess two assumptions that the bias estimator used in this work made: that we are performing unbiased estimates of bias parameters and that we have limited our measurements to scales where we do not observe strong running with the bias parameters. This is an important assumption, as in hybrid EFT, a running of the biases with $k_{\rm max}$ can be a sign of a potential breakdown of the bias expansion itself. This appendix concerns itself with validating these two assumptions. \par 
To validate that our estimator recovers ``unbiased" estimates of the bias parameters in full generality would require, for example, implementing separate-Universe simulations from which we can explicitly measure the LLIMD biases (as done in \cite{2016JCAP...02..018L}, which would require additional hydrodynamic and UniverseMachine simulations, beyond the scope of this work. \par
We thus opt, instead, to assess consistency between measurements of $b_1$ from the cross-correlation of the galaxy field with the dark matter field
\begin{equation}
\label{eq:Pgm_Pmm_estimate}
\begin{split}           
    b_1&= \lim_{k\to 0} \frac{\left\langle\delta_g(\textbf{k}) \delta_m\left(\textbf{k}^{\prime}\right)\right\rangle}{\left\langle\delta_m(\textbf{k}) \delta_m\left(\textbf{k}^{\prime}\right)\right\rangle}\\
    &=\lim_{k\to 0} \frac{P_{gm}}{P_{mm}},
\end{split}
\end{equation}
which is equivalent to the separate-universe definition of $b_1$.
\begin{figure}
    \centering
    \includegraphics[width=\columnwidth]{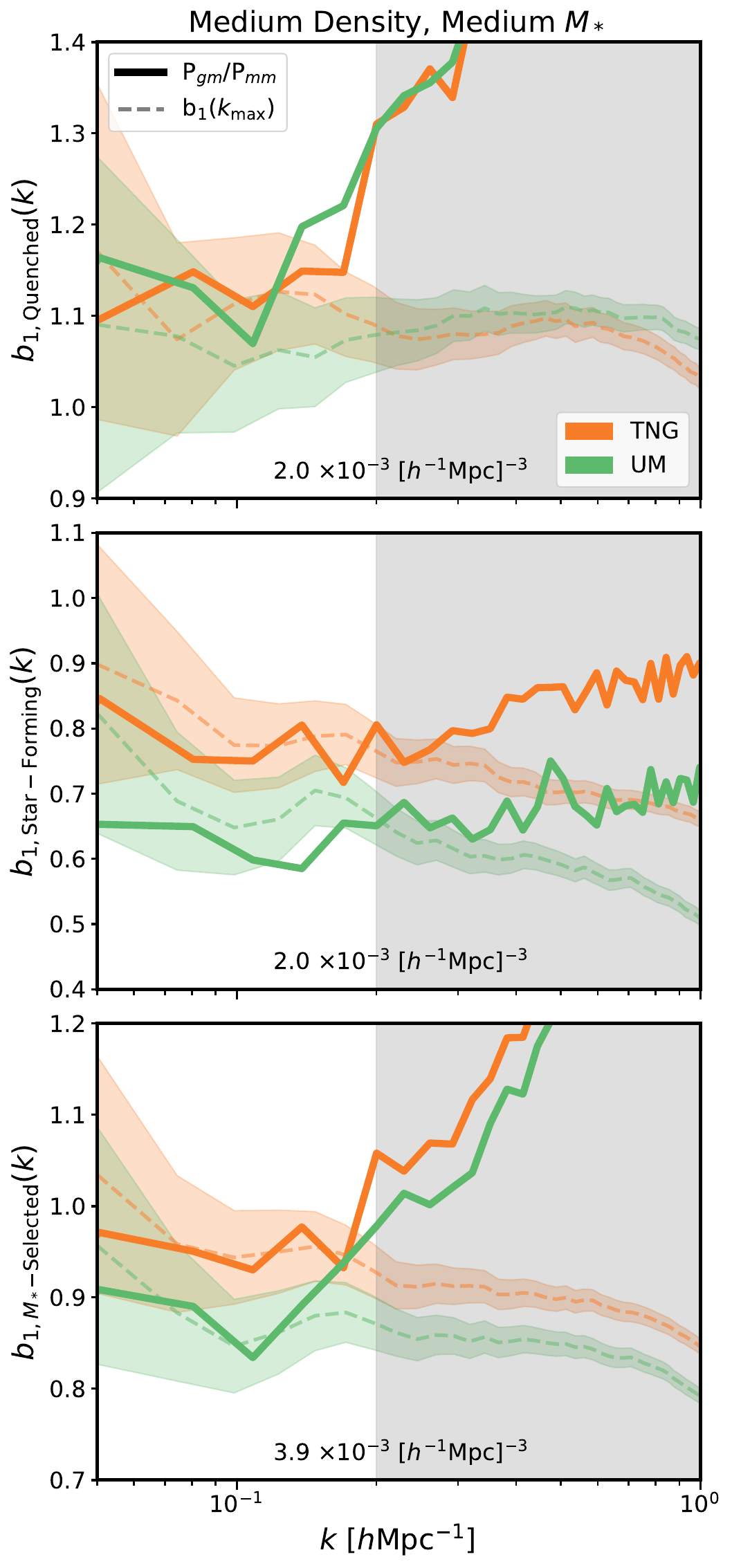}
    \caption{Consistency check of the large-scale approximation of the linear bias $b_1$ at $z=1$, computed via a ratio of the galaxy-matter cross-correlation power spectrum $P_{gm}$ to the matter-matter power spectrum $P_{mm}$. The approximation is plotted in solid lines for TNG (orange) and UM (green) as a function of scale $k$, while our measured $b_1(k_\mathrm{max})$ is plotted as a dashed line with shaded regions that show its theoretical error.
    The gray shaded region shows scales with $k_\mathrm{max} \ge 0.2\ihmpc$
    , which are discarded from our fit of the bias parameters.} 
    \label{fig:Pgm_Pmm}
\end{figure}
This comparison is shown, for the medium density sample at $z=1$ in  Figure \ref{fig:Pgm_Pmm}. This spectrum is representative of all other densities and redshifts considered. The cross-spectrum ratio is plotted in bold lines for quenched, star-forming, and $M_*$-selected galaxies with a medium number density cut at $z=1$ in TNG (orange) and UM (green). Our measured values of $b_1$ as a function of $k_\mathrm{max}$ are plotted as a dashed line, with the theoretical error bars plotted as the filled region. The gray shaded space to the right of each panel shows the $k_\mathrm{max}=0.2\ihmpc$ cut-off scale, beyond which includes modes that we do not include in our model.

Our measured values match the linear expectation of Equation~\ref{eq:Pgm_Pmm_estimate} at large scales. Second, we find that our measured values of linear bias $b_1(k_\mathrm{max})$ only begin running with scale relative to their uncertainties for $k_\mathrm{max}>0.2\ihmpc$. This check confirms that our estimates of $b_1$ are robust to at least $k_\mathrm{max} \sim 0.2 \ihmpc$, which is in contrast with the cross-spectrum estimates of $b_1$ that already deviate from a constant at $k > 0.1 \ihmpc$.
\begin{figure}
    \centering
    \includegraphics[width=\columnwidth]{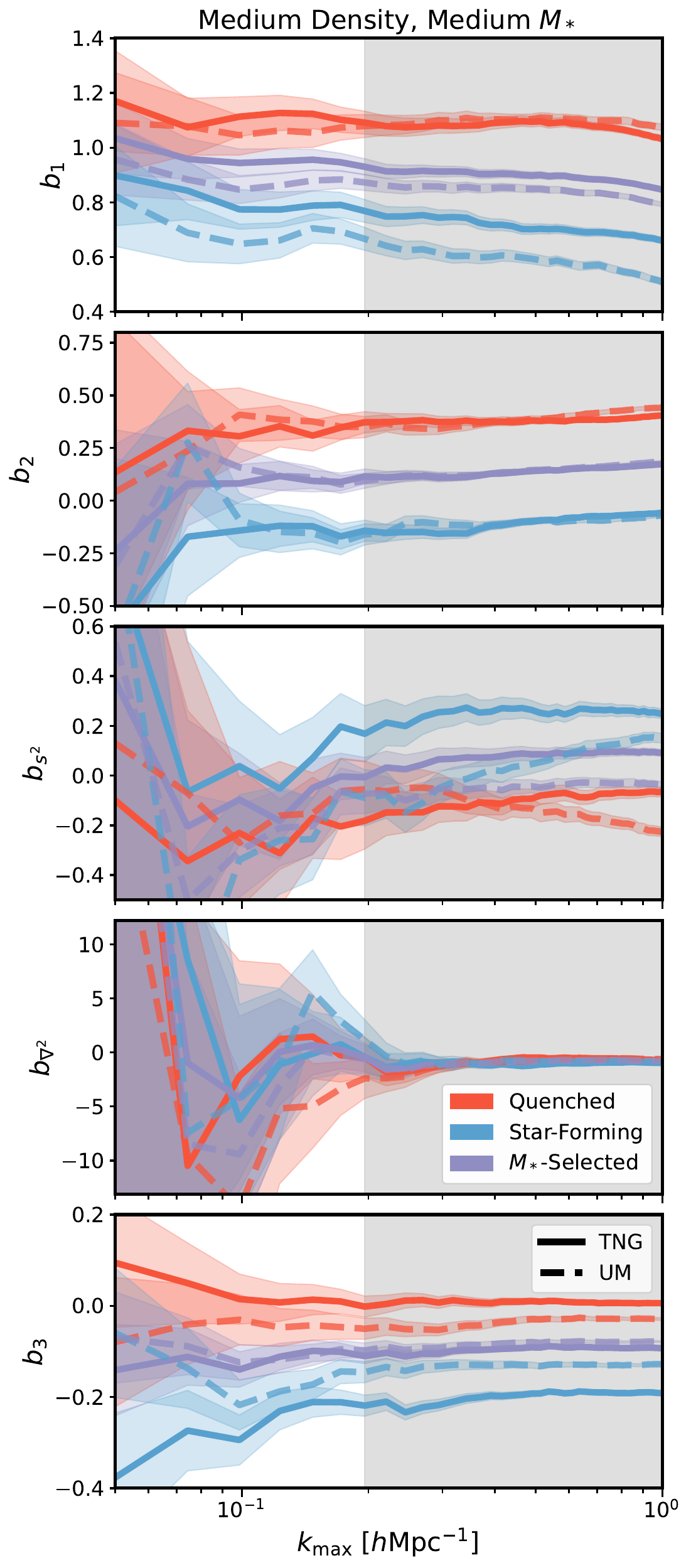}
    \caption{
    The bias parameter measurements $b_1$, $b_2$, $b_{s^2}$, $b_{\nabla^2}$, and $b_3$ at $z=1$ as a function of $k_\mathrm{max}$ for TNG (solid lines) and UM (dashed lines) for the quenched (red), star-forming (blue), and $M_*$-selected(purple) galaxy samples. Only the medium number density cut is shown. The theoretical error of the bias parameters is depicted by the shaded regions.} 
    \label{fig:b1b2}
\end{figure}

We also verify the lack of running up to $k_\mathrm{max}=0.2\ihmpc$ for all five bias parameters, as shown in Figure \ref{fig:b1b2}. Here, a medium number density cut is shown at $z=1$ for quenched (red), star-forming (blue), and $M_*$-selected (purple) galaxies in UM (solid lines) and TNG (dashed lines). Our measured values of the bias parameters \{$b_1$, $b_2$, $b_{s^2}$, $b_{\nabla^2}$, $b_3$\} are plotted as a function of $k_\mathrm{max}$, with the estimated uncertainties plotted as the filled region. The gray shaded space to the right of each panel shows the $k_\mathrm{max}=0.2\ihmpc$ cut-off scale, beyond which includes modes that we do not include in our model. We do note that the UM quenched and star-forming samples exhibit some large degree of running at scales beyond our cut-off, unlike the TNG measurements.

Like Figure \ref{fig:Pgm_Pmm}, we see in Figure \ref{fig:b1b2} that the bias parameters only begin running with scale for $k_\mathrm{max}>0.2\ihmpc$, showing that this cut-off scale is appropriate for measuring each of the bias parameters. This acts as a validation check of our approach to describing the field-level distribution of galaxies using HEFT to second order in the Lagrangian bias expansion. It is also interesting to note that the larger scatter in the linear bias $b_1$ for star-forming galaxies, as discussed in \S\ref{sec:galaxy model comparison}, is apparent in both Figures \ref{fig:Pgm_Pmm} and \ref{fig:b1b2}. This wider scatter is reflected in our priors for the linear bias of star-forming galaxies. As discussed in \S\ref{sec:priors_starforming} and shown in Figure \ref{fig:priors_b1}, the standard deviation of $b_1(z)$ is larger for the star-forming galaxies than the other galaxy types, at all redshifts we consider.

\section{Impact of galaxy type}
\label{app:galaxy type}

\begin{figure*}
    \centering  \includegraphics[width=\textwidth]{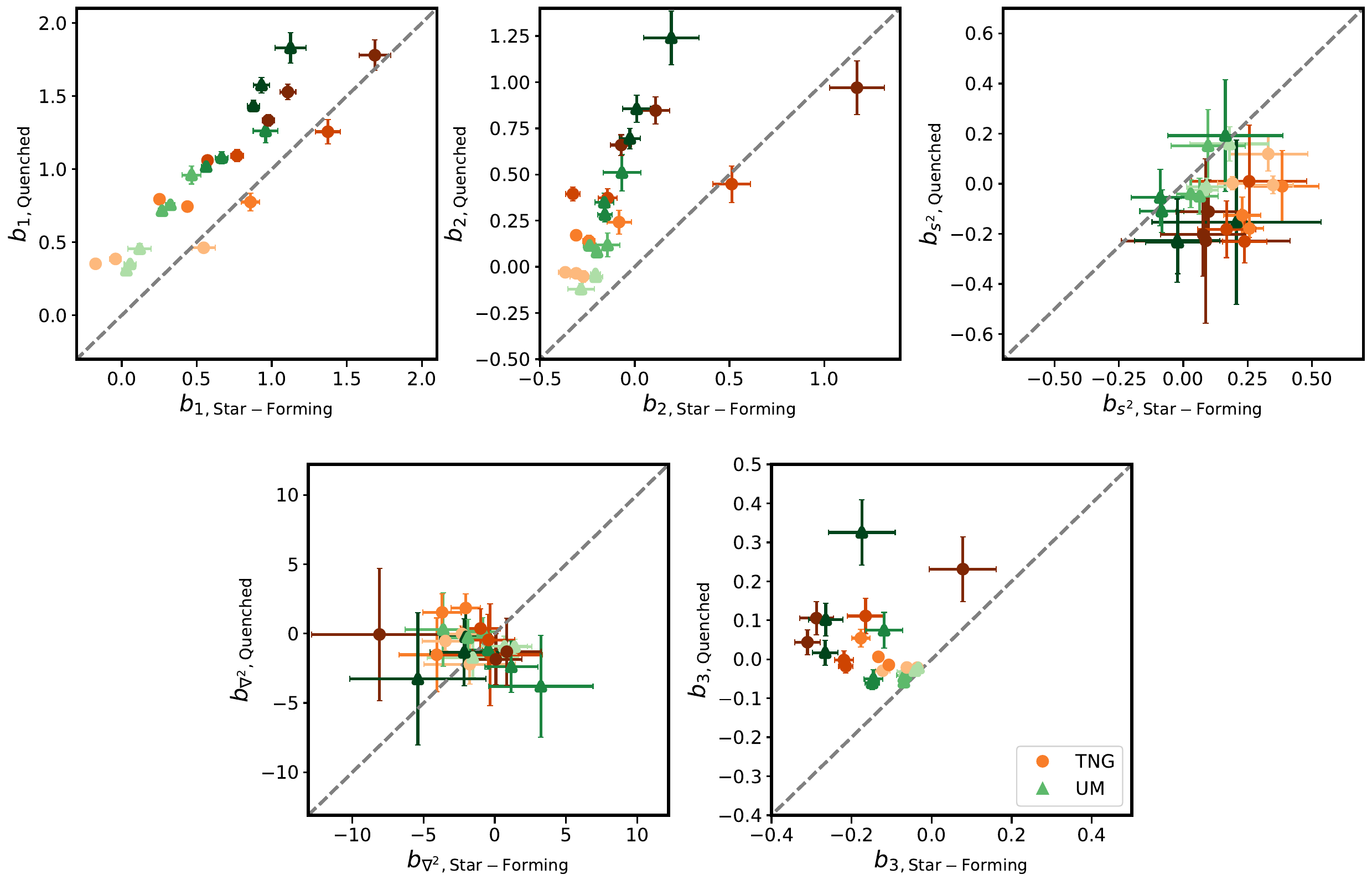}
    \caption{
    The bias parameter measurements of quenched galaxies in comparison to star-forming galaxies for samples at each number density and redshift ($z=\{0.0, 0.5, 1.0, 1.5\}$) in TNG (orange circles) and UM (green triangles). The color intensity (light to dark) maps to the redshift (low to high). 
    A 1:1 dashed line is plotted for reference.}
    \label{fig:quenched_starforming}
\end{figure*}

Expanding upon the discussion in \S\ref{sec:galaxy model comparison}, we further investigate the impact of galaxy type on the bias parameter measurements by comparing the quenched and star-forming populations directly in Figure \ref{fig:quenched_starforming}. We plot the bias measurements for matched quenched and star-forming samples at identical redshifts and number density cuts, in both UM (orange triangles) and TNG (gray circles). A 1:1 dashed line is plotted for reference.

Aside from some outliers in $b_1$ and $b_2$ for TNG, we see that none of the bias parameters follow this 1:1 relation between the two galaxy types. This supports our finding in \S\ref{sec:measurements} that quenched and star-forming galaxies occupy distinct parts of bias parameter space, with quenched galaxies being more heavily biased for the LIMD parameters $b_1$, $b_2$, and $b_3$. We find that the star-forming galaxies are more biased in $b_{s^2}$, a trend that is more pronounced in TNG than in UM. As discussed in \S\ref{sec:galaxy model comparison}, this is due to the higher value of $b_{s^2}$ for star-forming galaxies, and conversely lower value of $b_{s^2}$ for quenched galaxies, that we find in TNG compared to UM. In contrast, we find that there is no correlation in $b_{\nabla^2}$: both quenched and star-forming galaxies are centered around $b_{\nabla^2}\approx0$, although it appears that there is more spread in the star-forming galaxies. This increased scatter is reflected in our priors; as listed in Table \ref{tab:priors}, the standard deviation $\sigma_{b_\nabla^2}$ is largest for star-forming galaxies in comparison to the other galaxy types.

Finally, similar to our findings in \S\ref{sec:galaxy model comparison}, we find that the few outliers that follow the 1:1 relation in $b_1$ and $b_2$ belong to the lowest number density sample. This is consistent with the wider scatter in $b_1$ and $b_2$ that we see in Figure \ref{fig:bias_bias}. In Figure \ref{fig:b1b2}, however, we can determine that these low number density samples are from TNG, rather than UM. 
We thus find that TNG has a wider scatter in its measurements, and the largest discrepancies can be found in the highest-mass objects. 

\section{Full Priors}
\label{app:fullpriors}

\begin{figure*}
    \centering
    \includegraphics[width=\textwidth]{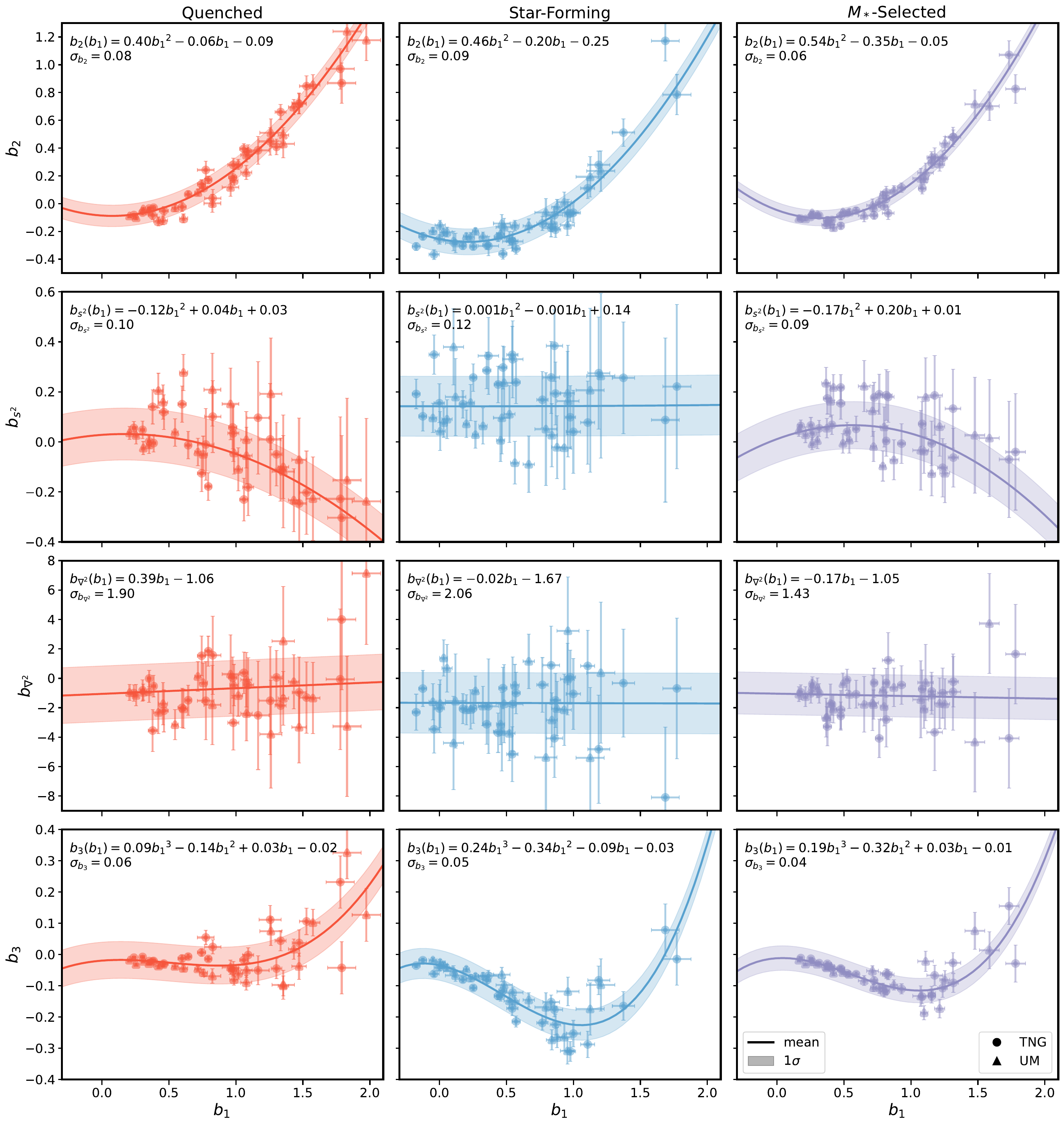}
    \caption{
    Gaussian priors for each of the bias parameter relations as a function of linear bias $b_1$. The bias parameter measurements are plotted for each sample of galaxies with and without AB in UM (triangles) and TNG (circles). Separate priors are set for the quenched (red), star-forming (blue), and the $M_*$-selected population (purple). The 1$\sigma$ standard deviation band is plotted around the best-fit polynomial, as labeled in each panel. This band characterizes the scatter between our range of galaxy formation prescriptions.}
    \label{fig:priors}
\end{figure*}

In a more detailed version of Figure \ref{fig:priors_bn}, we present our priors on the bias parameter relations $b_2(b_1)$, $b_{s^2}(b_1)$, $b_{\nabla^2}(b_1)$ and $b_3(b_1)$ in Figure \ref{fig:priors}. The best-fit curve is plotted as a solid line, and the shaded region depicts the standard deviation around this mean. The priors are identical to those shown in Figure \ref{fig:priors_bn}, but for readability, we have separated the quenched (red), star-forming (blue), and $M_*$-selected (purple) samples into different columns. For full transparency, we also elect to show all of the bias parameter measurements for which we make the best-fit curve. These are represented by circle and triangle markers for TNG and UM, respectively, and span all redshifts ($z=\{0.0, 0.5, 1.0, 1.5\}$) and number density cuts (high, medium, low). Both the original and AB-removed measurements are also included in each fit, for a total of 48 samples in each panel. This choice was made so that our analysis could span as full a range of galaxy formation physics as possible.

\begin{figure*}
    \centering
    \includegraphics[width=\textwidth]{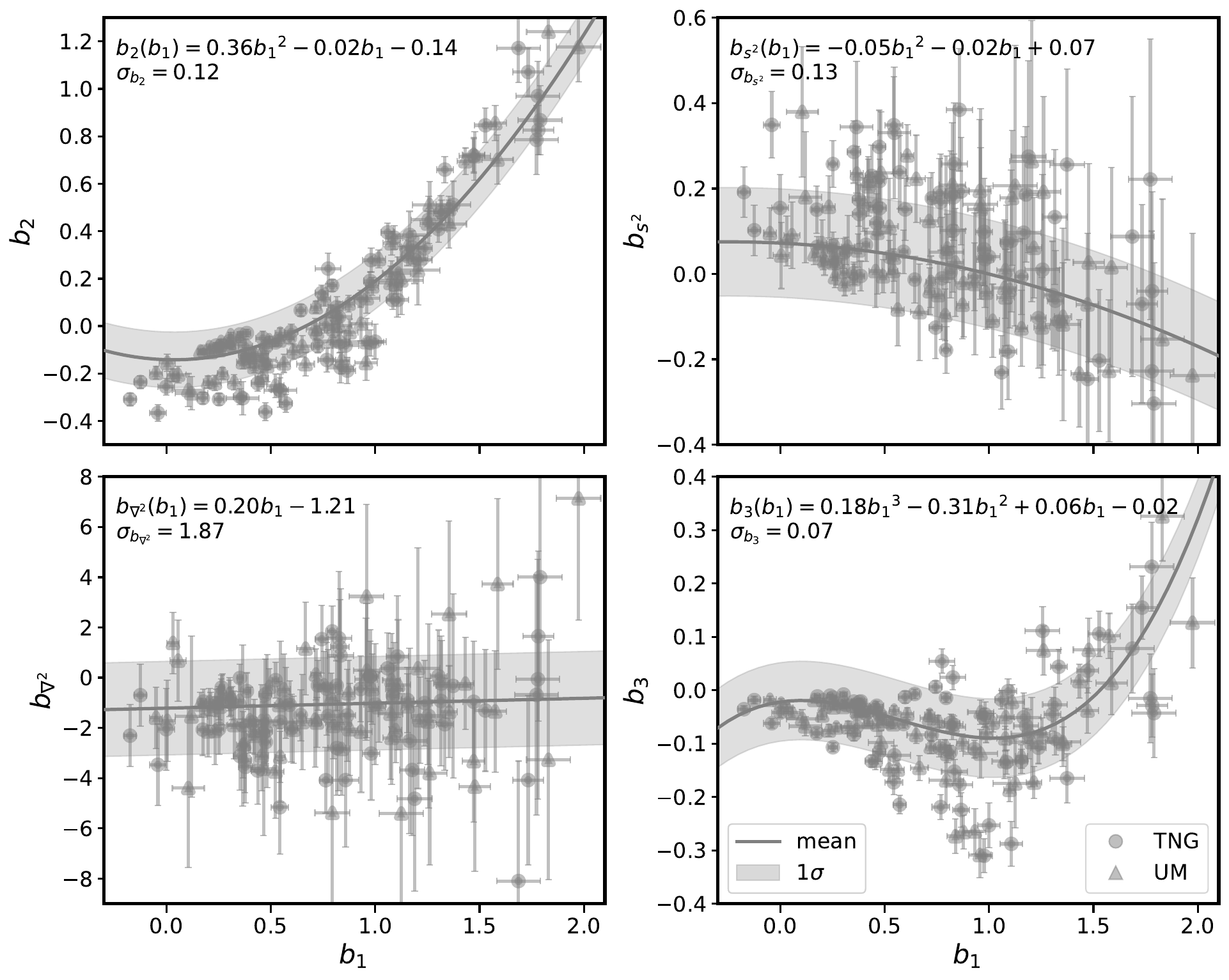}
    \caption{
    {Conservative Gaussian priors for each of the bias parameter relations as a function of linear bias $b_1$. The bias parameter measurements are plotted for each sample of galaxies with and without AB in UM (triangles) and TNG (circles). Priors are set for the all samples of galaxies together, including the quenched, star-forming, and the $M_*$-selected population. The 1$\sigma$ standard deviation band is plotted around the best-fit polynomial, as labeled in each panel.}
    }
    \label{fig:priors_all}
\end{figure*}

\begin{table}
    \centering
    \begin{tabular}{ll} 
    \midrule
    \midrule
     $b_i(b_1)$ & All Galaxies \\
     \midrule
    \multirow{2}{*}{$b_2(b_1)$} & $b_2=0.36b_1^2-0.02b_1-0.14$ \\
     & $\sigma_{b_2}=0.12$ \\
    \midrule
    \multirow{2}{*}{$b_{s^2}(b_1)$} & $b_{s^2}=-0.05b_1^2-0.02b_1+0.07$\\
    & $\sigma_{b_{s^2}}=0.09$ \\
    \midrule
     \multirow{2}{*}{$b_{\nabla^2}(b_1)$} & $b_{\nabla^2}=0.20b_1-1.21$\\
     & $\sigma_{b_{\nabla^2}}=1.43$ \\
    \midrule
    \multirow{2}{*}{$b_3(b_1)$} & $b_3=0.18b_1^3-0.31b_1^2+0.06b_1-0.02$\\    
    & $\sigma_{b_3}=0.04$\\
    \midrule
    \end{tabular}
    \caption{Conservative Gaussian priors on the bias parameter relations $b_2(b_1)$, $b_{s^2}(b_1)$, $b_{\nabla^2}(b_1)$, and $b_3(b_1)$ for all samples of galaxies together, including the quenched, star-forming, and $M_*$-selected population. Priors are defined by a best-fit line at $b_i(b_1)$ and the $1\sigma$ standard deviation.
    }
    \label{tab:priors_all}
\end{table}

\begin{figure}
    \centering
    \includegraphics[width=\columnwidth]{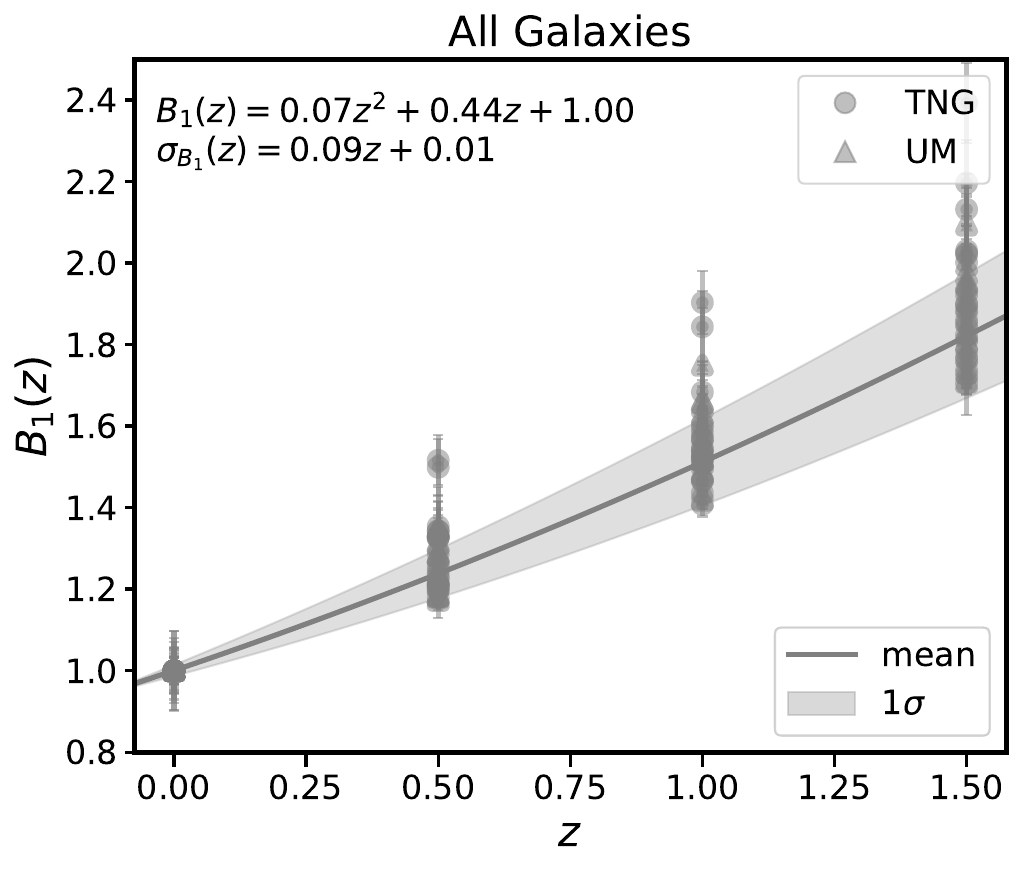}
    \caption{
    Conservative Gaussian priors for the time evolution of the linear bias $b_1$ relative to its value at $z=0$, parameterized as a fit to the mean 
    $B_1(z)=\frac{1+b_1(z)}{1+b_1(0)}$. The bias parameter measurements are plotted for each sample of galaxies with and without AB in UM (triangles) and TNG (circles). Priors are set for all of the samples of galaxies together, including the quenched, star-forming, and the $M_*$-selected population. The 1$\sigma$ standard deviation evolves linearly with redshift $z$ and is plotted as a shaded band around the best-fit polynomial, as labeled in each panel.}
    \label{fig:priors_b1_all}
\end{figure}

\begin{table}
    \centering
    \begin{tabular}{l} 
    \midrule
    \midrule
      All Galaxies \\
     \midrule
    $B_1(z)=0.07z^2+0.44z+1.00$\\
    $\sigma_{B_1}(z)=0.09z+0.01$ \\
    \midrule
    \end{tabular}
    \caption{Conservative Gaussian priors on the linear bias parameter $b_1$, parameterized via the best-fit mean to the function $B_1(z)=\frac{1+b_1(z)}{1+b_1(0)}$ for all samples of galaxies together, including the quenched, star-forming, and $M_*$-selected population. The $1\sigma$ standard deviation evolves linearly as a function of redshift $z$. 
    }
    \label{tab:priors_b1_all}
\end{table}

We also present in Figure \ref{fig:priors_all} an alternative, more conservative set of priors to those in Figures \ref{fig:priors_bn}, \ref{fig:priors_b1}, and \ref{fig:priors}. Instead of separating the samples by galaxy type, we treat all of the galaxy samples together and present one set of priors for all of the data, using the same method detailed in \S\ref{sec:priors}. We include these 144 data samples in the figure for comparison. The best-fit mean and $1\sigma$ standard deviation are listed in each panel of Figure \ref{fig:priors_all}, as well as in Table \ref{tab:priors_all}. We present the corresponding priors in $B_1(z)$ in Figure \ref{fig:priors_b1_all} and Table \ref{tab:priors_b1_all}. These priors may be more suitable for those desiring a less restrictive set of bounds in a HEFT-based cosmological analysis than the sample-specific priors presented in this work. 

\bibliography{main_revised}{}
\bibliographystyle{aasjournal}

\end{document}